%% file: ms.tex
\documentclass[10pt,journal,compsoc]{IEEEtran}
\usepackage{graphicx}
 \usepackage{adjustbox}
\usepackage{caption}
\usepackage{subcaption}
\usepackage{paralist}
\usepackage{tabularx}
\usepackage{colortbl}
\usepackage{balance}
\usepackage{booktabs}
\usepackage{multirow}
\usepackage{multicol}
\usepackage{pbox}
\usepackage{mathtools}
\usepackage{xcolor}
\usepackage[linesnumbered, ruled, vlined]{algorithm2e}
\usepackage{algorithmic}
\usepackage{paralist}
\usepackage{tabularx}
\usepackage{url}
\usepackage{balance}
\usepackage{multirow}
\usepackage{hyperref} 
\usepackage{multicol}
\usepackage{amsmath}
\usepackage{setspace}
\usepackage{verbatim}
\usepackage{float}
\usepackage[normalem]{ulem}
\usepackage{xspace}
\usepackage{amssymb}
\usepackage{ifthen}
\usepackage{pifont}
\usepackage{tikz}
\usepackage{anyfontsize}
\usepackage{hyperref}
\usepackage[most]{tcolorbox}
\definecolor{MidnightBlue}{HTML}{006895}

\newcommand{\ovts}{{\sc V2S}\xspace}

\newcommand{\vts}{{\sc V2S+}\xspace}

\newcommand{\vtss}{{\sc V2S+'s~}}

\newcommand{\ReDraw}{{\sc ReDraw}\xspace}

\newcommand{\approach}{\vts}

\newcommand{\Faster}{{\sc Faster R-CNN}\xspace}

\newcommand{\Rcnn}{{\sc R-CNN}\xspace}
\newcommand{\Cnn}{{\sc CNN}\xspace}
\newcommand{\Cnns}{{\sc CNNs}\xspace}
\newcommand{\AlexNet}{{\sc AlexNet}\xspace}

\newcommand{\VGGNet}{{\sc VGGNet}\xspace}
\newcommand{\numApps}{90\xspace}
\newcommand{\numVideos}{243\xspace}

\include{macro}

\ifCLASSOPTIONcompsoc
  \usepackage[nocompress]{cite}
\else
  \usepackage{cite}
\fi

\ifCLASSINFOpdf
\else
\fi

\begin{document}

\title{Translating Video Recordings of Complex Mobile App UI Gestures into Replayable Scenarios}

\author{
        Carlos Bernal-C\'ardenas,~\IEEEmembership{ Member,~IEEE,}
        Nathan Cooper,~\IEEEmembership{Student Member,~IEEE,}\\
        Madeleine Havranek, ~\IEEEmembership{Student Member,~IEEE,}
        Kevin~Moran,~\IEEEmembership{Member,~IEEE,}\\
		Oscar Chaparro,~\IEEEmembership{Member,~IEEE,}
        Denys Poshyvanyk,~\IEEEmembership{ Member,~IEEE,} and
        Andrian Marcus,~\IEEEmembership{Member,~IEEE}%

\IEEEcompsocitemizethanks{
\IEEEcompsocthanksitem{C. Bernal-C\'ardenas is with the Liquid Team, Microsoft, Redmond, WA, 98052.\protect\\
E-mail: carlosbe@microsoft.com}
\IEEEcompsocthanksitem{N. Cooper, M. Havranek, O. Chaparro, and D. Poshyvanyk are with the Department of Computer Science, College of William \& Mary, Williamsburg, VA, 23185.\protect\\
E-mail: [nacooper01, mrhavranek]@email.wm.edu,  oscarch@wm.edu,  denys@cs.wm.edu}
\IEEEcompsocthanksitem{K. Moran is with the Department
of Computer Science, George Mason University, Fairfax,
VA, 22030.\protect\\
E-mail: kpmoran@gmu.edu}
\IEEEcompsocthanksitem{A. Marcus is with the Department
of Computer Science, The University of Texas at Dallas, Richardson,
TX, 75080.\protect\\
E-mail: amarcus@utdallas.edu}}%
\thanks{Manuscript received Oct 2021;}}

\markboth{IEEE Transactions on Software Engineering,~Vol.~\#, No.~\#,~2022}%
{Bernal-C\'ardenas \etal: Translating Video Recordings of Mobile App UI Gestures into Replayable Scenarios for Native and Hybrid Apps}

\IEEEtitleabstractindextext{%
\begin{abstract}
Screen recordings of mobile applications are easy to obtain and capture a wealth of information pertinent to software developers (\eg bugs or feature requests), making them a popular mechanism for crowdsourced app feedback. Thus, these videos are becoming a common artifact that developers must manage.
In light of unique mobile development constraints, including swift release cycles and rapidly evolving platforms, automated techniques for analyzing all types of rich software artifacts provide benefit to mobile developers. Unfortunately, automatically analyzing screen recordings presents serious challenges, due to their graphical nature, compared to other types of (textual) artifacts.
To address these challenges, this paper introduces \vts, an automated approach for translating video recordings of Android app usages into replayable scenarios. \vts is based primarily on computer vision techniques and adapts recent solutions for object detection and image classification to detect and classify user \textit{gestures} captured in a video, and convert these into a replayable test scenario. \NEW{Given that \vts takes a computer vision-based approach, it is applicable to both hybrid and native Android applications. We performed an extensive evaluation of \vts involving \numVideos videos depicting  4,028 GUI-based actions collected from users exercising features and reproducing bugs from a collection of over \numApps popular native and hybrid Android apps.
Our results illustrate that \vts can accurately replay scenarios from screen recordings, and is capable of reproducing $\approx$ 90.2\% of sequential actions recorded in native application scenarios on physical devices, and $\approx$ 83\% of sequential actions recorded in hybrid application scenarios on  emulators, both with low overhead.
}
A case study with three industrial partners illustrates the potential usefulness of \vts from the viewpoint of developers.
\end{abstract}

}

\maketitle

\IEEEdisplaynontitleabstractindextext

\IEEEpeerreviewmaketitle
\input{1_intro.tex}

\input{2_background.tex}

\input{3_approach.tex}

\input{4_implementation.tex}

\input{5_study.tex}

\input{6_results.tex}

\input{7_related-work.tex}

\input{8_threats.tex}

\input{9_conclusion.tex}

\ifCLASSOPTIONcompsoc
  \section*{Acknowledgments}
\else
  \section*{Acknowledgment}
\fi

This work is supported in part by the NSF CCF-1927679, CCF-1815186, CNS-1815336, CCF-1955837, CCF-1910976, and CCF-1955853 grants. Any opinions, findings, and conclusions expressed herein are the authors' and do not necessarily reflect those of the sponsors.

\ifCLASSOPTIONcaptionsoff
  \newpage
\fi

\bibliography{minimum}
\bibliographystyle{IEEEtran}

\begin{IEEEbiography}
[{\includegraphics[width=1in,height=1.25in,clip,keepaspectratio]{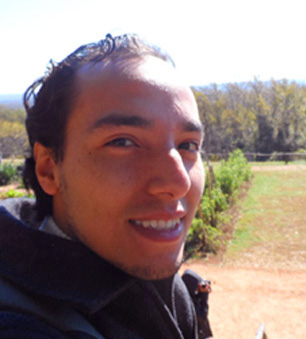}}]{Carlos Bernal-C\'ardenas} is currently a Software Development Engineer at Microsoft. He received the B.S. degree in systems engineering from the Universidad Nacional de Colombia in 2012 and his M.E. in Systems and Computing Engineering in 2015. He received a Ph.D. degree from William \& Mary in August 2021. His research interests include software engineering, software evolution and maintenance, information retrieval, software reuse, mining software repositories, mobile applications development, and user experience. He has published in several top peer-reviewed software engineering venues including: ICSE, ESEC/FSE, ICST, and MSR.  He has also received the ACM SIGSOFT Distinguished paper award at ESEC/FSE'15 \& '19 and ICSE'20. More information is available at \url{http://www.cs.wm.edu/~cebernal/ }.
\end{IEEEbiography}

\vspace{-4em}

\begin{IEEEbiography}[{\includegraphics[width=1in,height=1.25in,clip,keepaspectratio]{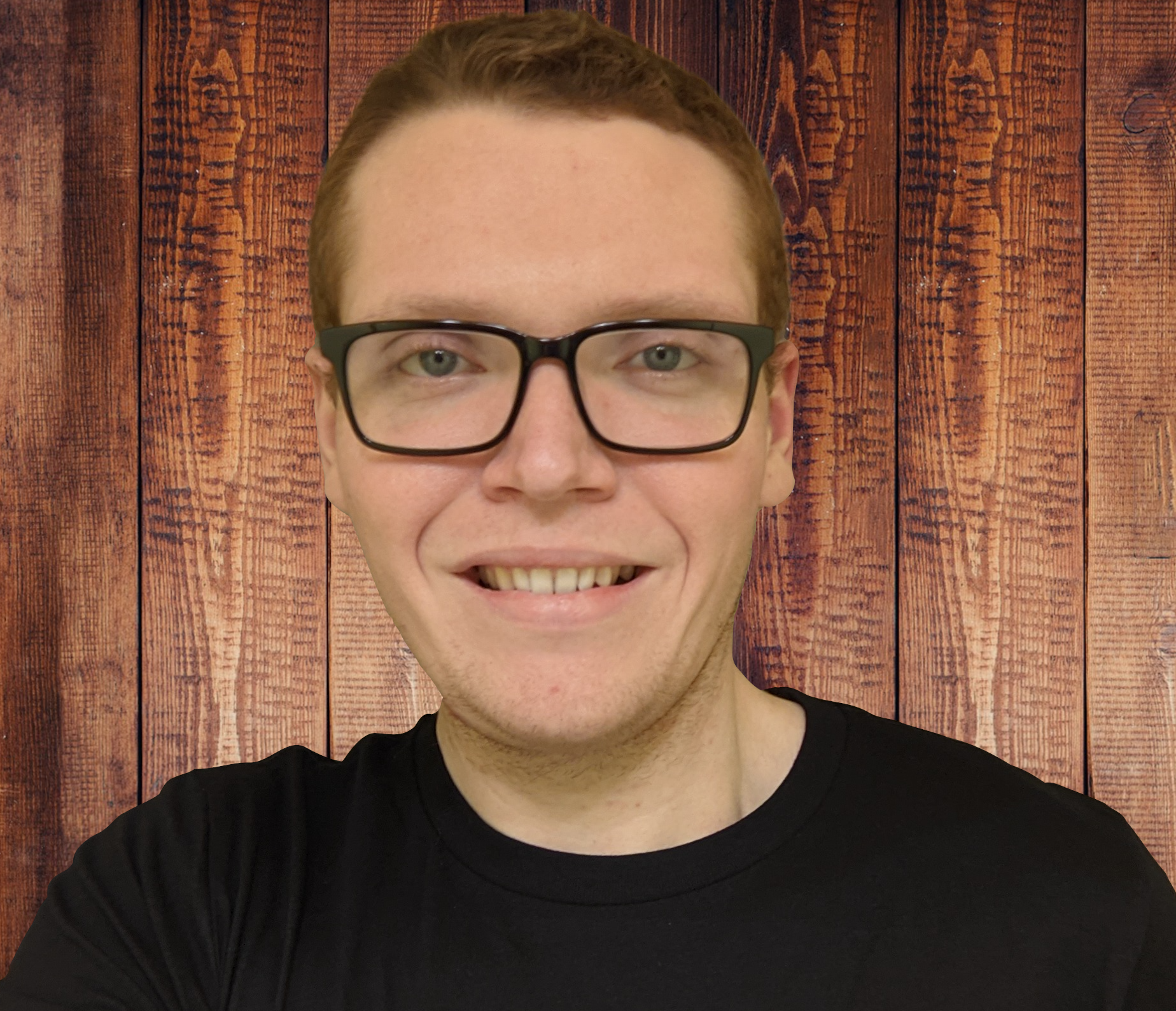}}]{Nathan Cooper} is a nerd and received a B.S. degree in Software Engineering from the University of West Florida in 2018. He is currently a Ph.D. candidate in Computer Science at William \& Mary under the mentorship of Dr. Denys Poshyvanyk and is a member of the Semeru Research group. He has research interests in Software Engineering, Machine / Deep Learning applications for Software Engineering, information retrieval, and question \& answering applications for Software Engineering. He has published in the top peer-reviewed Software Engineering venues ICSE and MSR. He has also received the ACM SIGSOFT Distinguished paper award at ICSE'20. More information is available at \url{https://nathancooper.io/#/}.
\end{IEEEbiography}

\vspace{-4em}

\begin{IEEEbiography}[{\includegraphics[width=1in,height=1.25in,clip,keepaspectratio]{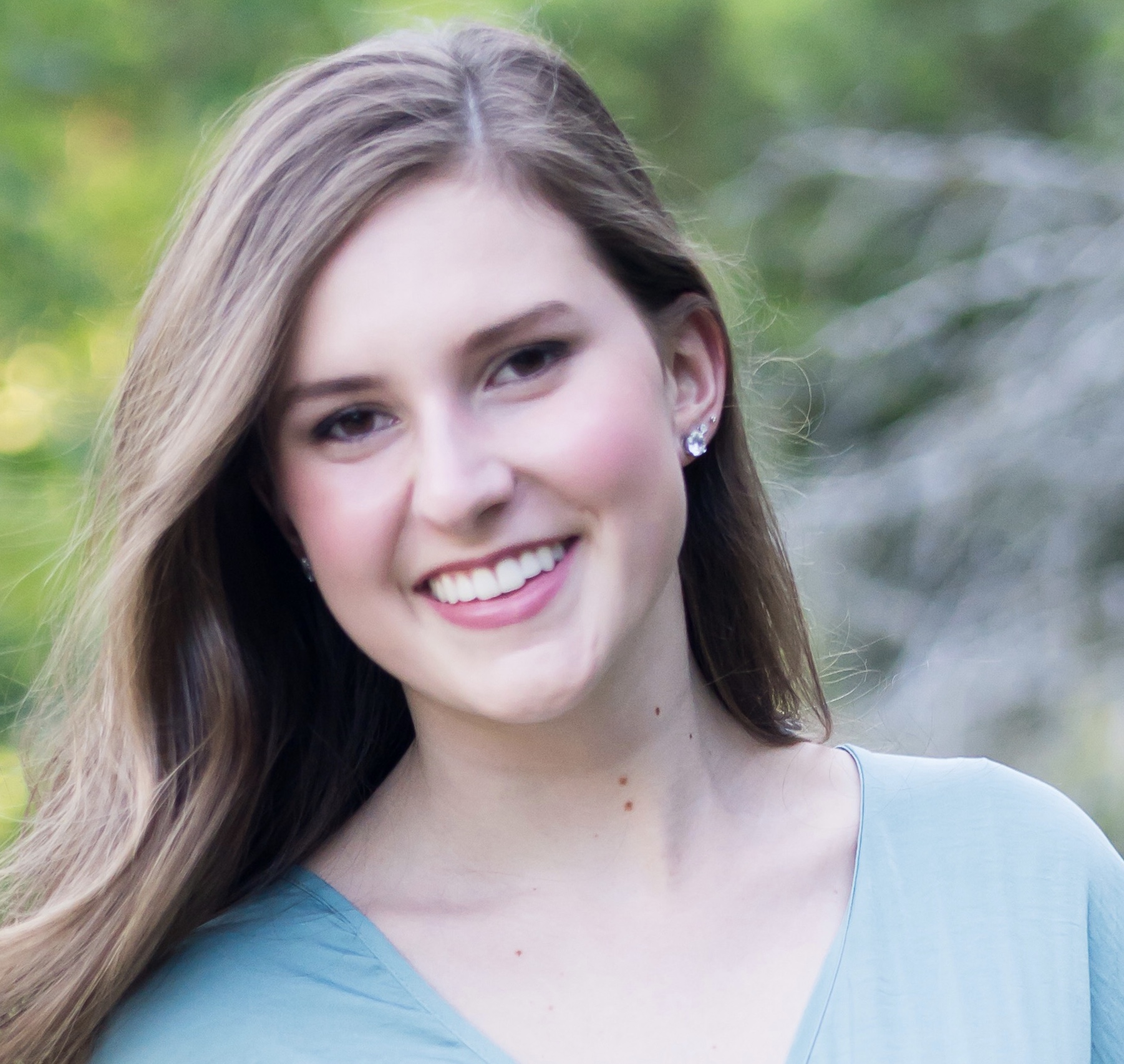}}]{Madeleine Havranek} is an undergraduate student in the Computer Science Department at William \& Mary. She is currently a member of the SEMERU research group and is pursuing an undergraduate honors thesis on the topic of automating software testing reproduction. Her research interests lie in applications of deep learning to software engineering and testing tasks. Havranek is an IEEE student member.
\end{IEEEbiography}

\vspace{-4em}

\begin{IEEEbiography}[{\includegraphics[width=1in,height=1.25in,clip,keepaspectratio]{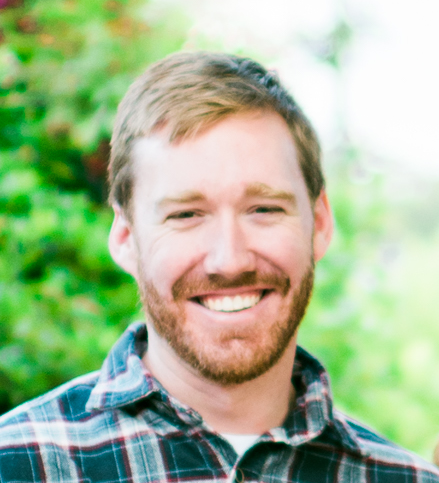}}]{Kevin Moran} is currently an Assistant Professor at George Mason University where he directs the SAGE research group. He graduated with a B.A. in Physics from the College of the Holy Cross in 2013 and an M.S. degree from William \& Mary in August of 2015.  He received a Ph.D. degree from William \& Mary in August 2018. His main research interest involves facilitating the processes of software engineering, maintenance, and evolution with a focus on mobile platforms. He has published in several top peer-reviewed software engineering venues including: ICSE, ESEC/FSE, TSE, USENIX, ICST, ICSME, and MSR. His research has been recognized with ACM SIGSOFT distinguished paper awards at ESEC/FSE 2019 and ICSE 2020, and a Best Paper Award at CODASPY'19. He is a member of the ACM and IEEE. More information is available at \url{http://www.kpmoran.com}
\end{IEEEbiography}

\vspace{-4em}

\begin{IEEEbiography}[{\includegraphics[width=1in,height=1.25in,clip,keepaspectratio]{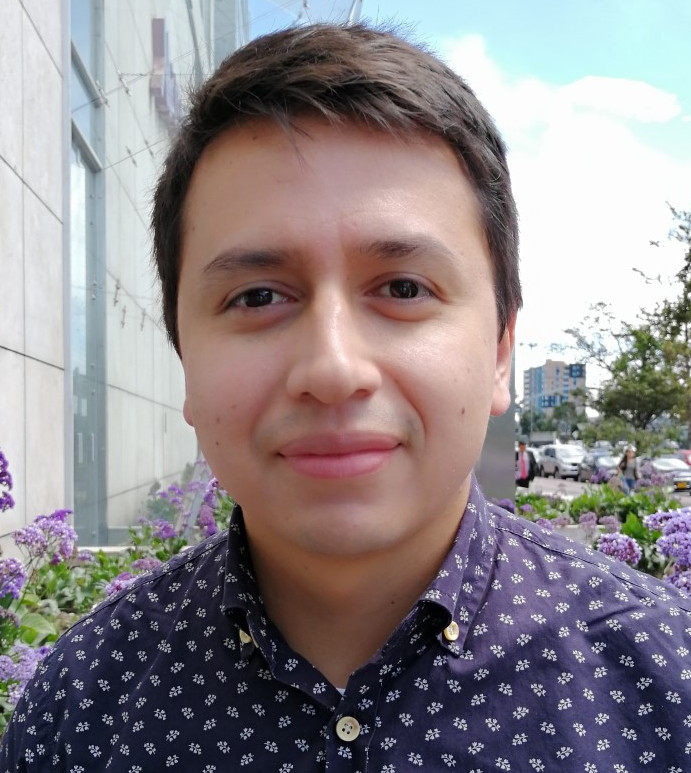}}]{Oscar Chaparro} is an Assistant Professor at William \& Mary. He received this Ph.D. (2019) in Software Engineering at the University of Texas at Dallas, advised by Dr. Andrian Marcus. He received his B.Eng. (2010) and M.Eng. (2013) degrees in Systems Engineering and Computing from the National University of Colombia. His research interests lie in software maintenance and evolution, program comprehension, code refactoring, code quality, developer’s productivity, and text analysis applied to software engineering. 
Oscar has authored several publications in top software engineering venues, such as ICSE, ESEC/FSE, EMSE, and ICSME. He received ACM SIGSOFT Distinguished Paper Awards at ESEC/FSE’19 and ICSE'20, and the IEEE TCSE Distinguished Paper Award at ICSME’17. Oscar has served on the organizing and program committees of several conferences and workshops, including ICSE, ASE, ICSME, MSR, ICPC, and the NLBSE'22, DySDoc3, and DocGen2 workshops. Oscar is a member of the IEEE and ACM. More information is available at: \url{https://ojcchar.github.io/}
\end{IEEEbiography}

\vspace{-4em}

\begin{IEEEbiography}[{\includegraphics[width=1in,height=1.25in,clip,keepaspectratio]{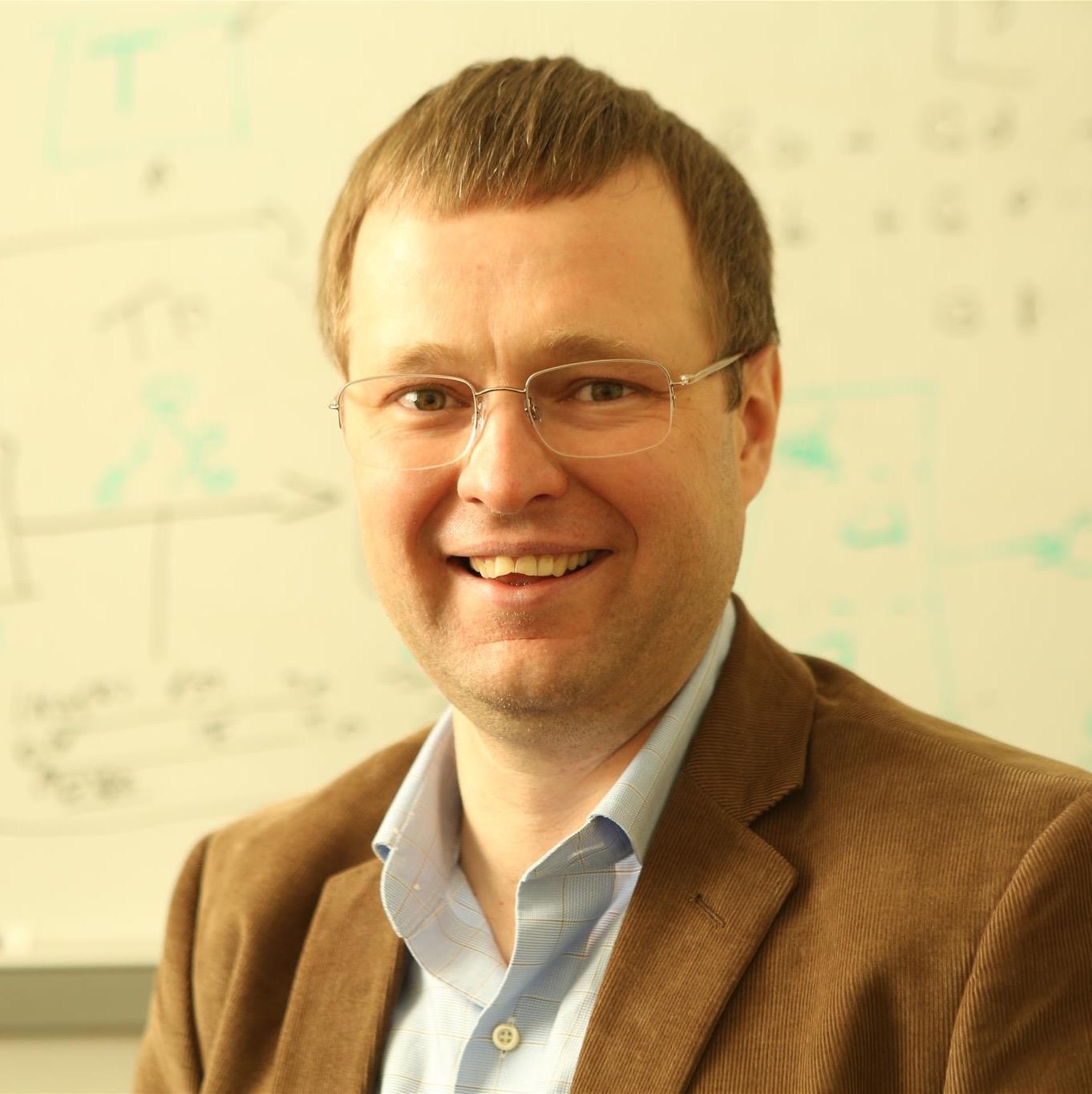}}]{Denys Poshyvanyk} is a Professor of Computer Science at William \& Mary. He received the MS and MA degrees in Computer Science from the National University of Kyiv-Mohyla Academy, Ukraine, and Wayne State University in 2003 and 2006, respectively. He received the PhD degree in Computer Science from Wayne State University in 2008. He served as a program co-chair for ASE'21, MobileSoft'19, ICSME'16, ICPC'13, WCRE'12 and WCRE'11. He currently serves on the editorial board of IEEE Transactions on Software Engineering (TSE), Empirical Software Engineering Journal (EMSE, Springer), Journal of Software: Evolution and Process (JSEP, Wiley) and Science of Computer Programming. His research interests include software engineering, software maintenance and evolution, program comprehension, reverse engineering and software repository mining. His research papers received several Best Paper Awards at ICPC'06, ICPC'07, ICSM'10, SCAM'10, ICSM'13, CODAPSY'19 and ACM SIGSOFT Distinguished Paper Awards at ASE'13, ICSE'15, ESEC/FSE'15, ICPC'16, ASE'17, ESEC/FSE'19 and ICSE'20. He also received the Most Influential Paper Awards at ICSME'16, ICPC'17, ICPC'20 and ICSME'21. He is a recipient of the NSF CAREER award (2013).  He is a senior member of the IEEE and ACM. More information is available at: \url{http://www.cs.wm.edu/~denys/}
\end{IEEEbiography}

\vspace{-4em}

\begin{IEEEbiography}[{\includegraphics[width=1in,height=1.25in,clip,keepaspectratio]{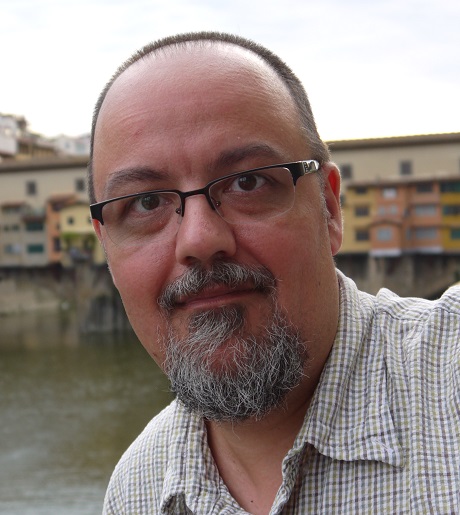}}]{Andrian Marcus} is a Professor of Computer Science and Software Engineering at the University of Texas at Dallas.
He obtained his Ph.D. in Computer Science from Kent State University (US), and has prior degrees in Computer Science and European Studies from The University of Memphis (US) and Babes-Bolyai University (Cluj-Napoca, Romania).
His research interests are in software engineering, with focus on program understanding and software evolution. 
He is a former Fulbright Scholar and the recipient of the NSF CAREER award.
The research with his students and collaborators earned six Best/Distinguished Paper Awards and six Most Influential Paper Awards at software engineering conferences.
His professional service includes serving on the Steering Committees of  ICSME and VISSOFT. He was the General Chair and the Program Co-chair of ICSME'11 ICSME'10, respectively, and Program Co-Chair for other conferences (ICPC'09, VISSOFT'13, SANER'17).
He served on the editorial boards of the Empirical Software Engineering Journal, the Journal of Software: Evolution and Process, and IEEE Transactions on Software Engineering.
More information is available at: \url{https://personal.utdallas.edu/~amarcus/}
\end{IEEEbiography}

\end{document}

%% file: 1_intro.tex
\vspace{-0.3cm}
\IEEEraisesectionheading{\section{Introduction}\label{sec:introduction}}
\label{sec:intro}

\IEEEPARstart{M}{obile} application developers rely on a diverse set of software artifacts to help them make informed decisions throughout the development process. These information sources include user reviews, crash reports, bug reports, and emails, among others. An increasingly common component of these software artifacts is graphical information, such as screenshots or screen recordings~\cite{Nayebi2020}. This is primarily due to the fact that they are relatively easy to collect and, due to the GUI-driven nature of mobile apps, they contain rich information that can easily demonstrate complex concepts, such as a bug or a feature request. 
In fact, many crowd-testing and bug reporting frameworks have built-in screen recording features to help developers collect mobile application usage data and faults~\cite{watchsend,testfairy, Instabug, birdeatsbug}. 
\textit{Screen recordings} that depict application usages are used by developers to: (i) help understand how users interact with apps~\cite{Schusteritsch:CHI'07,mrtappy}; (ii) process bug reports and feature requests from end-users~\cite{3Bettenburg:FSE08}; and (iii) aid in bug comprehension for testing related tasks~\cite{Mao:ASE17}. 
However, despite the growing prevalence of visual mobile development artifacts, developers must still manually inspect and interpret screenshots and videos in order to glean relevant information, which can be time consuming and ambiguous. %
The manual effort required by this comprehension process complicates a development workflow that is already constrained by language dichotomies~\cite{Moran:ICPC'18} and several challenges unique to mobile software, including: (i) pressure for frequent releases~\cite{Hu:ESYS14,Jones:2014}, (ii) rapidly evolving platforms and APIs~\cite{Linares:FSE13,Bavota:TSE15}, (iii) constant noisy feedback from users~\cite{Ciurumelea:SANER'17,DiSorbo:FSE'16,Palomba:ICSE17,Palomba:JSS'18,Palomba:ICSME15}, and (iv) fragmentation in the mobile device ecosystem~\cite{Han:WCRE'12,Wei:ASE'16,android-fragmentation} among others~\cite{Linares-Vasquez:ICSME'17B}. Automation for processing graphical software artifacts is necessary and would help developers shift their focus toward core development tasks.

\NEW{To improve and automate the analysis of video-related mobile development artifacts, we introduce \textit{Video to Scenario+} (\vts), an easy-to-use and modular automated approach for translating video screen recordings of Android app usages into replayable scenarios. 
We designed \vts to operate solely on a video file recorded from an Android device, and as such, it relies primarily on computer vision techniques. It is based on our previously-published work~\cite{Bernal:ICSE20}. %
\vts adapts recent Deep Learning (DL) models for object detection and image classification to accurately detect and classify different types of user actions performed on the screen. These classified actions are then translated into replayable scenarios that can automatically reproduce user interactions (from the video) on a target device, making \vts the first purely graphical Android record-and-replay technique.}

In addition to helping automatically process the graphical data that is already present in mobile development artifacts, \vts can also be used for improving or enhancing additional development tasks that do not currently take full advantage of screen-recordings, such as creating and maintaining automated GUI-based test suites, and crowd-sourcing functional and usability testing.

\NEW{We conducted a comprehensive evaluation of \vts  to assess its \textit{accuracy}, \textit{robustness}, \textit{efficiency}, and \textit{usefulness} in generating replayable scenarios of videos collected from users interacting with both native and hybrid applications. Users interacted with these applications using a variety of gestures, including both single- and multi-fingered gestures.}
\NEW{We evaluated \vts on a set of popular native Android applications on physical devices. 
Users reproduced bugs and exercised multiple features on the top-rated apps of 32 categories in the Google Play market using single-fingered gestures. We also evaluated \vts using videos collected from top-rated apps in 2 categories %
where users executed multi-fingered actions (\eg two-fingered rotations, two-fingered pinches to zoom, \etc).} \NEWEST{
We chose to conduct separate studies to evaluate \vtss performance on native and hybrid applications in order to assess its generalizability for these applications and highlight \vtss support on hybrid applications, given the relative lack of GUI testing tools available for these apps. Hybrid applications are built using a combination of web standards (\eg HTML and CSS) and native SDKs and target multiple platforms (\eg both iOS and Android), 
whereas native applications are built using only native SDKs and target a single platform~\cite{Malavolta}. 
Unlike other testing tools (\textit{e.g.,}  {\sc CRAFTDROID}~\cite{Lin} and {\sc Monkey}~\cite{monkey}), 
\vts does not require any GUI metadata and instead leverages the detected spatial location of each touch event on the screen, which can be useful for app testing regardless of the application framework (web and/or native).} 
\NEW{ For these hybrid applications, we analyzed \vtss performance with user scenarios recorded in 14 popular applications from 7 categories on emulated devices. We evaluated \vts on video recordings collected for all of these scenarios. We also assessed the overhead of our technique and %
conducted a case study with three industrial partners to understand the practical applicability of \vts.} 

The results of our evaluation indicate that \vts is \textit{accurate}, and is able to correctly reproduce $\approx$ 83\% - 90\% of events across collected videos. The approach is also \textit{robust} in that it is applicable to a wide range of popular native and hybrid apps currently available on Google Play. In terms of \textit{efficiency}, we found that \vts imposes acceptable overhead, and is perceived as \textit{potentially useful} by developers.\\

\noindent In summary, the main contributions of our work are:
\begin{itemize}
	\item{ \vts, the first record-and-replay approach for Android that functions purely on screen-recordings of app usages. \vts adapts computer vision solutions for object detection and image classification, to effectively recognize and classify single- and multi-fingered user actions in the video frames of a screen recording;}
	\item{An automated pipeline for dataset generation and model training to identify user interactions from screen recordings. \NEW{This pipeline is publicly available in our online appendix~\cite{appendix} and it is fully automated and documented for others to use and adapt.}} %
	\item{\NEW{The results of an extensive empirical evaluation of \vts that measures the \textit{accuracy}, \textit{robustness}, and \textit{efficiency} across \numVideos videos from over \numApps applications;}} %
	\item{The results of a case study with three industrial partners who develop commercial apps, highlighting \vtss potential usefulness, as well as areas for improvement and extension;} %
	\item{An online appendix~\cite{appendix} %
	, which contains examples of videos replayed by \vts, experimental data, source code, trained models, and our evaluation infrastructure to facilitate reproducibility of the approach and the evaluation results.}
\end{itemize}

\NEW{This paper is a substantial extension of our prior research published at the 42nd International Conference on Software Engineering (ICSE'20)\cite{Bernal:ICSE20}. The extension includes the following new contributions: (i) a new version of \ovts, called \vts, which is fully implemented in Python and is capable of recognizing and replaying single- and multi-fingered gestures, (ii) an expanded video data set that includes a new set of hybrid applications, as well as new mobile app usages which include multi-finger gestures; and (iii) two new research questions that focus on assessing the ability of \vts to replay videos recorded using both single- and multi-finger gestures on native and/or hybrid applications.}

%% file: 2_background.tex
\section{Background}
\label{sec:background}

We briefly discuss DL techniques for image classification and object detection that we adapt for touch/gesture recognition in \vts. %

\vspace{-0.2cm}
\subsection{Image Classification}
\label{subsec:back-CNNs}

Recently, DL techniques that make use of neural networks consisting of specialized layers have shown great promise in classifying diverse sets of images into specified categories. Advanced approaches leveraging Convolutional Neural Networks (CNNs) for highly precise image recognition~\cite{Krizhevsky:NIPS12, Simonyan:ICLR14,Zeiler:ECCV14,Szegedy:CVPR15,He:CVPR16} have reached human levels of accuracy for image classification tasks.

Typically, each \textit{layer} in a \Cnn performs some form of computational transformation to the data fed into the model. The initial layer usually receives an input image. This layer is typically followed by a convolutional layer that extracts features from the pixels of the image, by applying \textit{filters} (\aka kernels) of a predefined size, wherein the contents of the filter are transformed via a pair-wise matrix multiplication (\ie the convolution operation). Each filter is passed throughout the entire image using a fixed \textit{stride} as sliding window to extract \textit{feature maps}. Convolutional layers are used in conjunction with \textit{max pooling} layers to further reduce the dimensionality of the data passing through the network. The convolution operation is linear in nature. Since images are generally non-linear data sources, activation functions such as Rectified Linear Units (ReLUs) are typically used to introduce a degree of non-linearity. Finally, a fully-connected layer (or series of these layers) are used in conjunction with a \textit{Softmax classifier} to predict an image class. The training process for CNNs is usually done by updating the weights that connect the layers of the network using gradient descent and back-propagating error gradients.

\NEW{
Recently, new models that either improve upon \Cnn models or use other architectures have shown promise for image classification. EfficientNet \cite{tan2019efficient} used a separate neural network to learn to build an efficient and performant \Cnn architecture. The Transformer \cite{Vaswani2017transformer} architecture, which traditionally had been only used on natural language processing (NLP) tasks, has also been applied to image classification by using a novel tokenization scheme that allowed for an image to be converted into a sequence of visual tokens instead of subword tokens \cite{wu2020:vit}. Additionally, MultiLayer Perceptrons (MLPs) have also shown potential promise for image classification on par with state of the art models \cite{tolstikhin2021mixer}.

}

\vts implements a customized CNN for the specific task of classifying the opacity of an image segment to help identify GUI interactions represented by a touch indicator (see Section \ref{subsec:approach-detection}).

\subsection{Object Detection}
\label{subsec:back-RCNNs}

\begin{figure}[t]
    \begin{center}
		\includegraphics[width=\columnwidth]{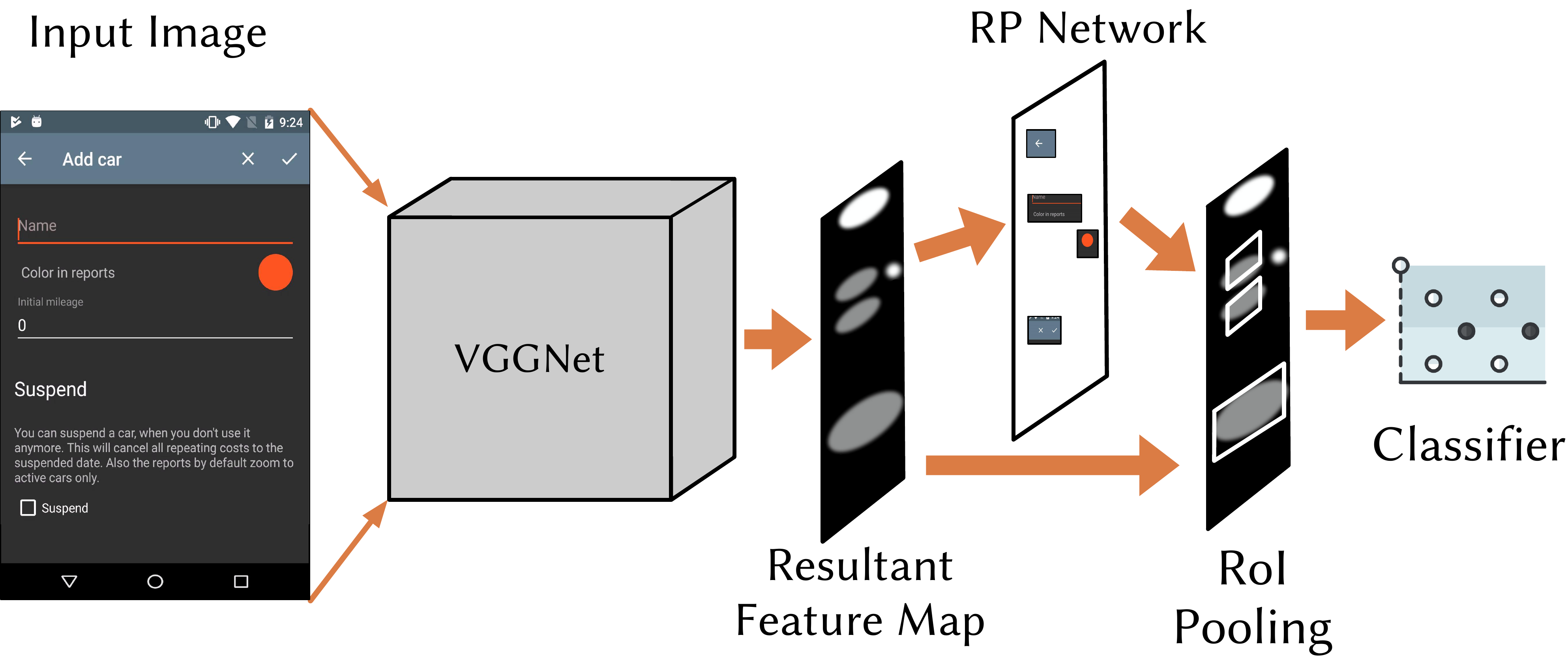}
		\vspace{-0.4cm}
        \caption{Illustration of the \Faster Architecture}
        \label{fig:faster-rcnn}
        \vspace{-0.8cm}
    \end{center}
\end{figure}

In the task of image classification, a single, usually more general label (\eg \textit{bird} or \textit{person}) is assigned to an entire image. However, images are typically multi-compositional, containing different objects to be identified. %

Similar to image classification, DL models for object detection have advanced dramatically in recent years, enabling object tracking and counting, as well as face and pose detection among other applications. One of the most influential neural architectures that has enabled such advancements is the Region-based CNN (\Rcnn) introduced by Girshick \etal~\cite{Girshick:CVPR14}. The \Rcnn architecture combines algorithms for image region proposals (RPs), which aim to identify image regions where content of interest is likely to reside, with the classification prowess of a \Cnn. An \Rcnn generates a set of RP bounding-boxes using a selective search algorithm~\cite{Uijlings:IJCV'13}. Then, all identified image regions are fed through a pre-trained \AlexNet~\cite{Krizhevsky:NIPS12} (\ie the \textit{extractor}) to extract image features into vectors. These vectors are fed into a support vector machine (\ie the \textit{classifier}) that determines whether or not each image region contains a class of interest. Finally, a greedy non-maximum suppression algorithm (\ie the \textit{regressor}) is used to select the non-overlapping regions with the highest likelihood as classified objects. \NEW{Recently, DETR \cite{carion2020detr} was proposed as an end-to-end way of performing object detection using a \Cnn and an encoder-decoder Transformer architecture. DETR does not need many of the hand-craft heuristics that the \Rcnn architecture requires making the architecture simpler and the authors found it performs both in accuracy and speed on par with the hyper optimizer \Faster~\cite{Ren:NIPS15} architecture.
}

We leverage the \Faster architecture~(Fig. \ref{fig:faster-rcnn}), which improves upon \Rcnn architecture through the introduction of a separate neural network to predict image region proposals. 
By integrating the training of the region proposal network into the end-to-end training of the network, both the speed and accuracy of the model increase. For \vts, we adapt the \Faster model to detect a touch indicator representing a user action in video frames (see Section \ref{subsubsec:approach-faster-rcnn}).

%% file: 3_approach.tex
\section{The V2S Approach}
\label{sec:approach}

\begin{figure*}[t]
    \begin{center}
		\includegraphics[width=0.93\linewidth]{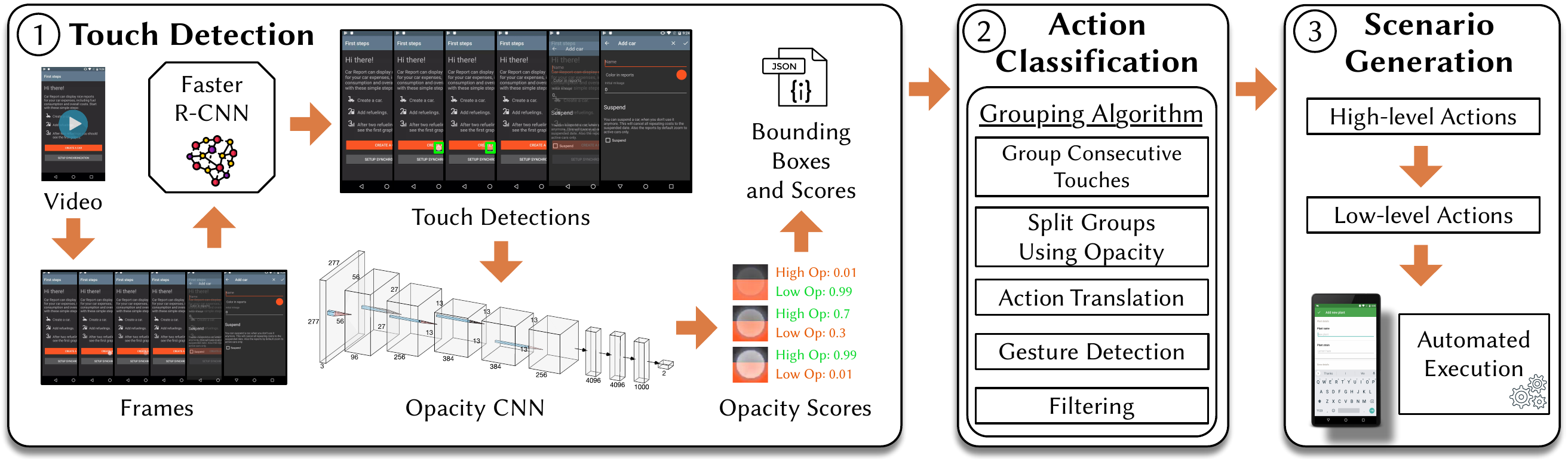}
		\vspace{-0.3cm}
        \caption{The \vts approach and phases (numbered)}
        \label{fig:approach}
        \vspace{-0.2cm}
    \end{center}
\end{figure*}

This section outlines the \vts approach for automatically translating Android screen recordings into replayable scenarios. %
\figref{fig:approach} depicts \vtss execution flow, which is divided into three main phases: (i) the \textit{Touch Detection} phase, which identifies user touches in each frame of an input video; (ii) the \textit{Action Classification} phase that groups and classifies the detected touches into discrete user actions (\ie \texttt{Tap}, \texttt{Long Tap}, \texttt{Gesture}, and \texttt{Multi-Fingered Gestures})%
, and (iii) the \textit{Scenario Generation} phase that exports and formats these actions into a replayable script. %

{\ovts \cite{Bernal:ICSE20} was originally implemented in Java, Bash, and Python. We have since redesigned the \vts structure to be implemented entirely in Python with extension and modularity in mind.  Each phase and component of \vts can now be easily substituted and/or extended by way of abstract classes which makes the entire pipeline malleable to fit a wide variety of research and development contexts.} 
Before discussing each phase in detail, we discuss some preliminary aspects of our approach, input specifications, and requirements.

\subsection{Input Video Specifications}
\label{subsec:approach-prelim}

\begin{figure}[t]
    \begin{center}
		\includegraphics[width=0.6\columnwidth]{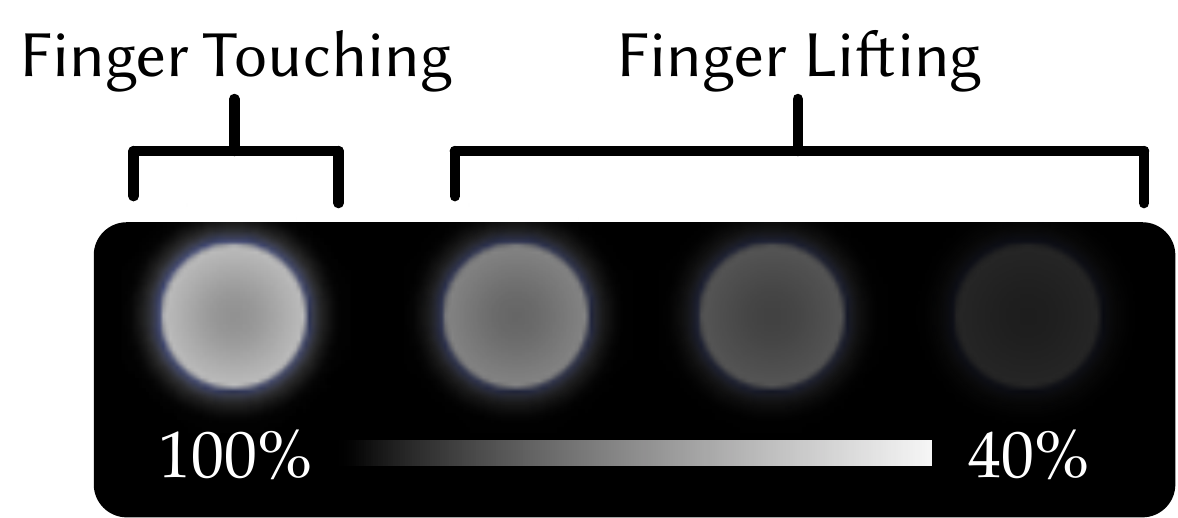}
		\vspace{-0.2cm}
        \caption{Illustration of touch indicator opacity levels}
        \label{fig:touch-opacity}
        \vspace{-0.7cm}
    \end{center}
\end{figure}

	In order for a video to be consumable by \vts, it must meet a few requirements to ensure proper functioning with our computer vision (CV) models. First, the video frame size must match the full-resolution screen size of the target Android device, in order to be compatible with a specified pre-trained object-detection network. This requirement is met by nearly every modern Android device that has shipped within the last few years. These videos can be recorded either by the built-in Android \texttt{screenrecord} utility, or via third-party applications~\cite{g-play-recording-apps}. The second requirement is that input videos must be recorded with at least 30 ``frames per second'' (FPS), which again, is met or exceeded by a majority of modern Android devices. This requirement is due to the fact that the frame-rate directly corresponds to the accuracy with which ``quick'' gestures (\eg fast scrolling) can be physically resolved in constituent video frames. \NEWR{ Third, the video needs to start at the initial screen/state of the application once it is opened or at the home screen of the device. This will guarantee that the reproduction will not miss any actions that are required to reproduce the video.} Finally, the videos must be recorded with the ``Show Touches'' option enabled on the device, which is accessed through an advanced settings menu~\cite{show-touches}, and is available by default on nearly all Android devices since at least Android 4.1.  
	This option renders a \textit{touch indicator}, which is a small semi-transparent circle, that gives a visual feedback when the user presses her finger on the device screen. The opacity of the indicator is fully solid from the moment the user first touches the screen and then fades from more to less opaque when a finger is lifted off the screen (see \figref{fig:touch-opacity}).

\subsection{Phase 1: Touch Detection}
\label{subsec:approach-detection}

The \textit{goal} of this phase is to accurately identify the locations where a user touched the device screen during a video recording.  To accomplish this, \vts leverages the DL techniques outlined in Sec.~\ref{sec:background} to both accurately find the position of the touch indicator appearing in video frames, and identify its opacity to determine whether a user's finger is being pressed or lifted from the screen. The main reason to use DL techniques over traditional image processing techniques (\eg Canny edge detection \cite{canny} and color analysis) is that the latter do not perform consistently on video frames
with varying levels of contrast (\eg between the touch indicator and the background), as they require different parameter values, depending on the image. 
\NEWEST{For example, canny edge detection requires specifying two thresholds or parameters 
(\ie min and max value for the gradient) that help to identify the edges in an image. However, finding the optimal values for these parameters for all the images across the videos is challenging because there can be different levels of contrast in the images across all the videos, due to the variety of the visual content that each application can show during its execution.
At the start of our research project, we experimented with a set of parameter values; however, we quickly realized that finding the optimal parameters would demand a large effort (\ie a wide range of experiments with different values) without guaranteeing the selected parameters would generalize to other images and apps. 
By using Deep Learning techniques, we account for the variability in contrast across the images from video screen recordings of mobile apps, controlling for experimental effort and improving generalizability.
} 

More specifically, we adapt an implementation of \Faster~\cite{Ren:NIPS15,Ren:PAMI17}, which makes use of \VGGNet~\cite{Simonyan:ICLR14} for feature extraction of RPs in order to perform touch indicator detection  (see Fig. \ref{fig:faster-rcnn}). To differentiate between low and high-opacity detected touch indicators, we build an {\sc Opacity CNN}, which is a modified version of \AlexNet~\cite{Krizhevsky:NIPS12}. Given that we adapt well-known DL architectures, here we focus on describing our adaptions, and provide model specs in our appendix~\cite{appendix}. %

The \textit{Touch Detection} Phase begins by accepting as input a video that meets the specifications outlined in \secref{subsec:approach-prelim}. First, the video is \textit{parsed} and decomposed into its constituent frames. Then, the \Faster network is utilized to \textit{detect} the presence of the touch indicator, if any, in every frame. Finally, the {\sc Opacity CNN} \textit{classifies} each detected touch indicator as having either low or high-opacity. The output of this phase is a structured \texttt{JSON} with a set of touch indicator bounding boxes in individual video frames wherein each detected touch indicator is classified based on the opacity.

\subsubsection{Parsing Videos}
\label{subsubsec:appraoch-parsing}

Before \vts parses the video to extract single frames, it must first normalize the frame-rate for those videos where it may be variable to ensure a constant FPS. Certain Android devices may record variable frame-rate video for efficiency~\cite{stackoverflow-android-screen-record}. This may lead to inconsistencies in the time between frames, which \vts utilizes in the action classification and scenario replay phases to synthesize the timing of touch actions. Thus, to avoid this issue, we normalize the frame rate to 30fps and extract individual frames using the \texttt{FFmpeg} tool~\cite{ffmpeg}.  %

\subsubsection{Faster R-CNN}
\label{subsubsec:approach-faster-rcnn}

After the individual frames have been parsed from the input video, \vts applies its object detection network to localize the bounding boxes of touch indicators.  %
However, before using the object detection, it must be trained. As described in \secref{sec:background}, the DL models that we utilize typically require large, manually labeled datasets to be effective. However, to avoid the manual curation of data, and  make the \vts approach practical, we designed a fully automated dataset generation and training process. %
To bootstrap the generation of \vtss object detection training dataset, we make use of the existing large-scale \ReDraw dataset of Android screenshots~\cite{Moran:TSE18}. This dataset includes over 14k screens extracted from the most popular Android applications on Google Play using a fully-automated execution technique. 

To create the dataset for training these individual models, we randomly sample 5k unique screenshots of different apps and programmatically superimpose an image of the \textit{touch indicator} at a random location in each screenshot. During this process, we took two steps to ensure that our synthesized dataset reflects actual usage of the touch indicator: (i) we varied the opacity of the indicator icon between 40\% and 100\% to ensure our model is trained to detect instances where a finger is lifted off the screen; (ii) we placed indicator icons on the edges of the screen to capture instances where the indicator may be occluded. This process is repeated three times per screenshot to generate 15k unique images. We then split this dataset using a 70\%-30\% random partitioning to create the training and testing sets respectively. We performed this partitioning such that all screenshots except one appear only in the testing set, wherein the one screenshot that overlapped had a different location and opacity value for the touch indicator. During testing, we found that a training set of 15k screens was large enough to train the model to extremely high levels of accuracy (\ie $>$ 97\%).

To train these models we use the \textit{TensorFlow Object Detection API}~\cite{TFODA} that provides functions and configurations of well-known DL architectures. We provide details regarding our training process for \vtss object detection network in \secref{subsec:study-rq1}. Note that, despite the training procedure being completely automated, it needs to be run only once for a given device screen size, after which it can be re-used for inference. After the model is trained, inference is run on each frame, resulting in a set of output \textit{bounding box} predictions for each screen, with a confidence score.

\subsubsection{Opacity CNN}
Once \vts has localized the screen touches that exist in each video frame, it must then determine the opacity of each detected touch indicator to aid in the \textit{Action Classification} phase. This helps \vts in accurately identifying instances where there are multiple actions in consecutive frames with very similar locations \NEWEST{or when a swipe ends, which is especially important for fast flicks due to Android needing to calculate the momentum of the swipe}. \NEWR{For example, if a user is typing on the keyboard and quickly taps the same key multiple times} \NEWEST{or double taps a back button, it causes to have multiple touch indicators with similar locations. Thus, identifying when an action finishes and another starts, by using touch indicator's opacity, facilitates action separation and accurate recognition of gestures that involve single and multiple fingers.}  %

To differentiate between low and high-opacity touch indicators, \vts adopts a modified version of the \AlexNet \cite{Krizhevsky:NIPS12} architecture as an {\sc Opacity CNN} that predicts whether a cropped image of the touch indicator is fully opaque (\ie finger touching screen) or has low opacity (\ie indicating a finger being lifted off the screen). 

Similar to the object detection network, we fully automate the generation of the training dataset and training process for practicality. We again make use of the \ReDraw dataset and randomly select 10k unique screenshots, randomly crop a region of the screenshot to the size of a touch indicator, and an equal number of full and partial opacity examples are generated. %
For the low-opacity examples, we varied the transparency levels between 20\% and 80\% to increase the diversity of samples in the training set. During initial experiments, we found that our {\sc Opacity CNN} required fewer training samples than the object detection network to achieve a high accuracy (\ie $>$ 97\%). Similar to the \Faster model, this is a one-time training process, however, this model can be re-used across varying screen dimensions. Finally, \vts runs the classification for all the detected touch indicators found in the previous step. Then  \vts generates a \texttt{JSON} file containing all the detected bounding boxes, confidence levels, and opacity classifications. %

\subsection{Phase 2: Action Classification}
\label{subsec:approach-classification}

The \texttt{JSON} file generated by the \textit{Touch Detection} phase contains detailed data about the bounding boxes, opacity information, and the video frame corresponding to each detected touch indicator\footnote{We use ``touch indicator'' and ``touch'' interchangeably moving forward.}. This \texttt{JSON} file is used as input into the \textit{Action Classification} phase where single touches are grouped and classified as high-level actions (\eg \texttt{Taps}). 

\NEW{The classification of these actions involves three main parts: 
(i) a \textit{grouping algorithm} that associates touches across subsequent frames into groups, which represent single discrete actions; 
(ii) \textit{action translation}, which identifies the action type of the grouped touches; and
(iii) \textit{single/multi-fingered action identification}, which identifies which of the detected actions are single- and multi-fingered. 
The result of the \textit{Action Classification} phase is a structured list of actions (\ie \texttt{Tap}, \texttt{Long Tap}, \texttt{Gesture}, or \texttt{Multi-Fingered Gesture}), where each action is a series of touches, or a series of actions in the case of Multi-Fingered Gestures, associated with video frames and screen locations.} %

\begin{figure}[t]
    \begin{center}
		\includegraphics[width=0.8\columnwidth]{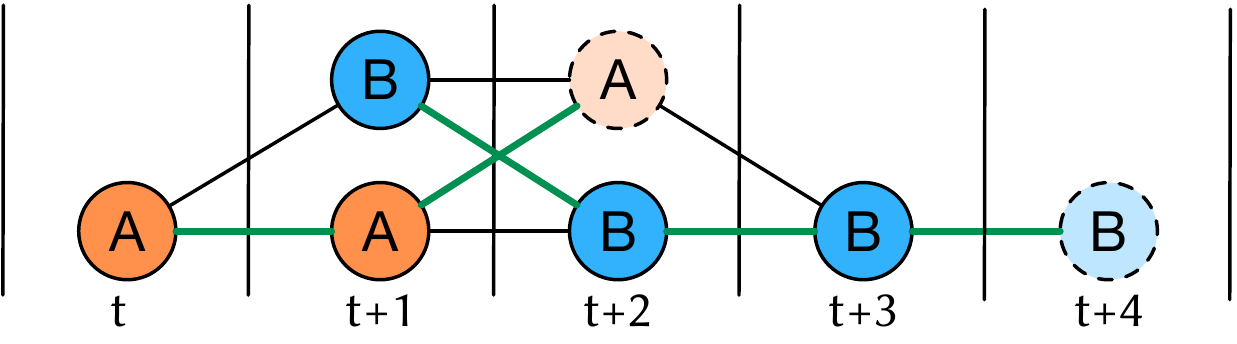}
		\vspace{-0.3cm}
        \caption{Illustration of the graph traversal problem for splitting discrete actions. Faded nodes with dotted lines represent touches where a finger is being lifted off the screen.}
        \label{fig:split_groups}
        \vspace{-0.6cm}
    \end{center}
\end{figure}

\subsubsection{Action Grouping}
\label{subsubsec:approach-grouping}

The first step of \vtss \textit{action grouping} step filters out detected touches where the model's confidence is lower than 0.7. The second step groups touches belonging to the same atomic action according to a tailored heuristic and a graph connection algorithm. This procedure is necessary because discrete actions performed on the screen will persist across several frames, and thus, need to be grouped and segmented accordingly. %

\noindent\textbf{Grouping Consecutive Touches.} The first heuristic groups touch indicators present in consecutive frames into the same group. As a measure taken to avoid (the rare occurrence of) a falsely detected touch indicator, touches that exist across two or fewer frames are discarded. This is due to the fact that, we observed in practice, even the quickest of touchscreen taps last across at least five frames.

\noindent\textbf{Discrete Action Segmentation.} There may exist successive single-fingered touches that were carried out extremely fast, such that there is no empty frame between two touches. In other cases, the touch indicator of one action may not fully disappear before the user starts a new action, leading to two or more touch indicators appearing in the same frame. These two situations are common when a user is swiping through a list and quickly tapping an option from the list, or typing quickly on the keyboard. Or, in the case of multi-fingered actions, two or more touches may appear in the same frame as part of the same gesture, but each component finger's action must be classified separately. \vts is able to determine where one action ends and another action begins as follows. 

\vts analyzes groups of consecutive or overlapping touches and segments them into discrete actions per finger using a heuristic-based approach. We model the grouping of touch indicators as a graph connectivity problem (see  \figref{fig:split_groups}). In this formulation, touch indicators are represented as nodes, vertical lines separate consecutive frames, and edges are possible connections between consecutive touches that make up a discrete action. The goal is to derive the proper edges for traversing the graph such that all touches for action $A$ are in one group and all touches for action $B$ are in another group (illustrated by the green edges in \figref{fig:split_groups}). Our algorithm decomposes the lists of consecutive touches grouped together into a graph. Starting from the first node, our algorithm visits each subsequent node and attempts to link it to the previous node. If there is only one node in a subsequent frame, then the two successive nodes are linked. If there is more than one node in a subsequent frame, our algorithm looks at the spatial distance between the previous node and both subsequent nodes, and groups previous nodes to their closer neighbors (as shown between frame $t$ and $t+1$). However, if multiple nodes in one frame are at a similar distance from the previous node (as between frames $t+1$ and $t+2$ in \figref{fig:split_groups}), then the opacity of the nodes is used to perform the connection. For example, it is clear that node $A$ in frame $t+2$ is a finger being lifted off the screen. Thus, it must be connected to the previously occurring action $A$ (\ie in $t+1$), and not the action $B$ that just started.%

Finally, after this process, the opacity of all linked nodes are analyzed to determine further splits. %
Distinct actions with no empty frame between them may exist. Therefore, if a low-opacity node is detected in a sequence of successively connected nodes, they are split into distinct groups representing different actions.

\NEW{At the end of this stage, distinct groups of touches that represent atomic actions (\eg \texttt{Tap}, \texttt{Long Tap}, or \texttt{Gesture}) are identified.}

\subsubsection{Action Translation}
\label{subsubsec:approach-translation}

\NEW{%
This process analyzes the derived groups of touches 
and classifies them based on the duration and touch locations of each group. %
Actions in which all of the touch indicators remain inside a specified small radius on the screen are classified as a \texttt{Tap} or a \texttt{Long Tap}. %
To correctly determine the threshold for this radius, we take direction from the Android Open Source Project, which defines the touch slop (\ie the distance in pixels that a touch can "wander`` before it is no longer interpreted as a tap) to be 8px~\cite{touch-slop1, touch-slop2, touch-slop3}. 
The action is classified as a \texttt{Tap} if it lasts across 20 or fewer frames, and as a \texttt{Long Tap} otherwise. Every other action is classified as a \texttt{Gesture}. %
}%

\noindent\textbf{Filtering.}
\vts removes actions that have touch indicators with low average opacity (\eg $<$ 0.1\%) across a group, as this could represent a rare \textit{series} of misclassified touch indicators from the \Faster. \vts also removes groups whose size is below or equals a threshold of two frames, as these might also indicate rare misclassifications. 

\NEW{After this stage, actions have been detected and classified, where each action is represented as a frame group with touch locations and a type (\ie \texttt{Tap},  \texttt{Long Tap}, or \texttt{Gesture}).}

\NEW{
\begin{algorithm}[t]
\DontPrintSemicolon 
\caption{\footnotesize{ Single/Multi-Fingered Action Identification}}
\label{alg:mfa-group}
\SetAlgoLined
\footnotesize
\KwIn{Detected Actions (DA)} 
\KwOut{Single- and Multi-Fingered Actions (SFAs, MFAs)}
\BlankLine
$DAc = countMultiTouchFrames(DA)$ \\
$pMFAs = select(DA, DAc > 0.5)$ \tcp{Potential MFAs}
$SFAs = select(DA, DAc \leq 0.5)$ \tcp{Initial SFAs}

\BlankLine
\BlankLine

$SDA = sortActions(pMFAs)$ \tcp{By starting frame}
$S = createNewStack()$ \tcp{Entries are action groups}
$initializeStack(S, SDA_1)$ \tcp{Pushes the 1st action}

\BlankLine
\BlankLine
\tcp{For each action}
\ForEach{$Action\ A \in \textit{SDA}$} {

    $AG = popActionGroup(S) $\\
    \tcp{For each action in the group}
    \ForEach{$Action\ at \in AG$} {
    
    \If {checkOverlap(A, at)}{
        $addToActionGroup(AG,A)$ \\
        \textbf{break}}
    }
    \If{notInActionGroup(A,AG)}{
    $PushNewActionGroup(S, A)$}
}

\BlankLine
\BlankLine

$MFAs = []$  \tcp{List of MFAs}
\ForEach{$Action\ Group\ AG \in \textit{S}$} {
    $c = countActions(AG)$

    \tcp{Does the group have a single action?}
    \eIf {$c == 1$}{  
        $addSFA(SFAs, AG)$
    }
    {
        $addMFA(MFAs, AG)$\\
    }
    
}

\end{algorithm}}

\subsubsection{Single/Multi-Fingered Action Identification}
\NEW{It is crucial that \vts is able to detect and classify both \textit{SFAs} and \textit{MFAs}. Many applications rely on \textit{MFAs} to accomplish core functionality within the app (\eg Google Maps utilizes two-finger rotations to allow users to rotate the map on the screen). If \vts cannot support these \textit{MFAs}, then its utility is limited to only certain applications and gestures, which is undesirable.}

\NEW{\vts first identifies frame groups that contain  \textit{Single-Fingered Actions} (SFAs) and \textit{Multi-Fingered Actions} (MFAs). 

\textit{MFAs} are collections of \textit{SFAs} that occur within the same frames, with each SFA corresponding to a single finger. After each action has been identified through the above steps (\ie described in Sect. \ref{subsubsec:approach-grouping} and \ref{subsubsec:approach-translation}), \vts regroups these actions based on their overlapping frames so that \textit{SFAs} and \textit{MFAs} can be classified appropriately and translated correctly into low-level events to be executed on the device (see Sect. \ref{subsec:approach-generation}). \vts executes Algorithm \ref{alg:mfa-group} to do so.

Initially, frame groups where more than 50\% of the frames have multiple detected touches are considered \textit{potential MFAs} (lines 1-2 in Algorithm \ref{alg:mfa-group}). Other frame groups are considered as \textit{SFAs} (line 3).} \NEW{Based on our initial experiments and observations, true \textit{MFAs} have a percentage of frames with multiple detected touches much higher than 50\% and \textit{SFAs} have a much smaller percentage than 50\%. Thus, 50\% is  a reasonable value for such a  threshold. Furthermore, when a user is tapping quickly on the keyboard, touch indicators may appear on the screen at the same time as one another, yet each finger's actions still must be identified as \textit{SFAs} rather than \textit{MFAs}. By introducing this 50\% threshold, we reduce the likelihood that these gestures will be linked together because it is unlikely that they overlap for over 50\% of their duration.}%

\vts then sorts the \textit{potential MFAs} by their starting frame in \textit{sortActions} (line 4), and creates a stack ($S$) which holds groups of actions (\ie \textit{MFAs}), by calling \textit{createNewStack} (line 5). Each group on the stack is a list of actions that execute simultaneously during the duration of the action group (on a device). %
The group on the top of the stack is the one currently being built. \vts initializes the stack (via \textit{initializeStack}) by pushing the first chronological detected action onto the stack to begin the first group (line 6).

Next, \vts considers each of the sorted detected actions (line 7: $A \in SDA$) and each of the actions within the group being built on the top of the stack (lines 8-9: $at \in AG$). \vts uses the \textit{checkOverlap} function to determine whether both actions $A$ and $at$ occur within common frames: if the first frame of $A$ occurs before the last frame of $at$, the actions overlap and \vts appends $A$ to the current action group $AG$ (lines 10-13). 
For example, consider the action $A$=\texttt{Gesture:$[$3,...,20$]$} with \texttt{$[$3,...,20$]$} being the list of frames where the gesture occurs. For a group $AG$=\texttt{$[$Gesture$_a$:$[$1,...,20$]$}, \texttt{Gesture$_b$:$[$2,...,17$]]$}, Algorithm \ref{alg:mfa-group} will append $A$ to $AG$ because the first frame of $A$ (\ie frame \#3) occurs before the last frame of \texttt{Gesture$_a$} (\ie frame \#20), as indicated by the frame indices (frame \#$3 <$ frame \#$20$).

If \vts has compared every action $at \in AG$ to $A$, and none of these actions overlap (line 15), \vts finishes building $AG$, and pushes $A$ on the stack $S$, starting a new action group. Continuing with the above example, if $AG$ is now \texttt{$[$Gesture$_a$:$[$1,...,20$]$}, \texttt{Gesture$_b$:$[$2,...,17$]$}, \texttt{Gesture$_c$:$[$3,...,20$]]$} and $A$ is \texttt{Gesture:$[$70,...,92$]$}, \vts will not append $A$ to $AG$ because the starting frame of $A$ occurs after the ending frame of any action $at \in AG$ (\ie frame \#70 $>$ frame \#20). Therefore, \vts pushes $A$ onto $S$ and $A$ becomes the first action of new group being built in $S$. Since \vts first sorts the detected actions by their starting frames, it is safe to close $AG$ when a given $A$ does not belong to $AG$ as any subsequent action will not overlap either. 

After all of the actions have been processed, \vts iterates over the created action groups (line 20). If the group length is one (line 22), the group contains a single action and is considered as a \textit{SFA} (line 23). Otherwise, \vts considers the group as a \textit{MFA} (line 25).

\textbf{Multi-Fingered Action Classification.} Once \vts has successfully identified \textit{SFAs} and \textit{MFAs}, \textit{MFAs} are further classified according to the number of fingers involved in the actions. This is computed by calculating the number of touch indicators per frame that is most frequent across all the frames in a \textit{MFA}. For example, if the majority of the frames within a \textit{MFA} contain three touch indicators, the MFA is classified as a \texttt{Three-Fingered Gesture}. 

We used this approach because we observed that, in rare instances, \vts incorrectly classifies application imagery similar to the touch indicator as an additional \texttt{Tap} on the screen. When executing the \textit{MFA}, this small \texttt{Tap} does not usually derail the successful execution of the scenario, so we do not remove it from the \textit{MFA}. However, if \vts only relied on the number of actions detected within an \textit{MFA} during the classification phase, the pipeline will then incorrectly  classify the number of fingers in the \textit{MFA}. For example, \vts would classify the \texttt{Three-Fingered Gesture} referenced above as a \texttt{Four-Fingered Gesture} because of this additional \texttt{Tap}, which harms \vtss classification accuracy. Instead, because most actions that compose an \textit{MFA} start and end during the same frames and the additional \texttt{Tap} usually lasts a shorter duration, classification of the number of fingers for the \textit{MFA} is more accurately represented by calculating the number of touch indicators per frame that is most frequent across all the frames in a~\textit{MFA}.  %

\vts supports classification of one- to ten-fingered gestures in an attempt to allow support for actions involving all fingers of a user’s hands. We recognize that, in practice, higher-digit gestures occur infrequently, and thus in \vtss evaluation (see Sect. \ref{sec:study}), we mainly focused on two-fingered \textit{MFAs} (\eg pinch to zoom or rotations) when considering \vtss support for multi-touches.

\subsection{Phase 3: Scenario Generation}
\label{subsec:approach-generation}

After all the actions have been derived by the \textit{Action Classification} phase, \vts proceeds by generating commands using the Android Debug Bridge (\texttt{adb}) that replays the classified actions on a device. To accomplish this, \vts converts the classified, high-level actions into low-level instructions in the \texttt{sendevent} command format, which is a utility included in Android's Linux kernel. Then, \vts uses a modified RERAN~\cite{Gomez:ICSE13} binary to replay the events on a device.%

\noindent\textbf{Generating the Scenario Script.} The \texttt{sendevent} command uses a  limited instruction set in order to control the UI of an Android device. The main instructions of interest are the \texttt{start\_event}, \texttt{end\_event}, ($x$, $y$) coordinates where the user's finger touched the screen, and certain special instructions required by devices with older Android API levels. %
\NEW{To create the script, low-level actions are generated in chronological order, but \textit{SFAs} and \textit{MFAs} are translated using separate techniques because their translation procedures progress slightly differently. %
}

\begin{figure}[ht]
\begin{center}
		\includegraphics[width=\linewidth]{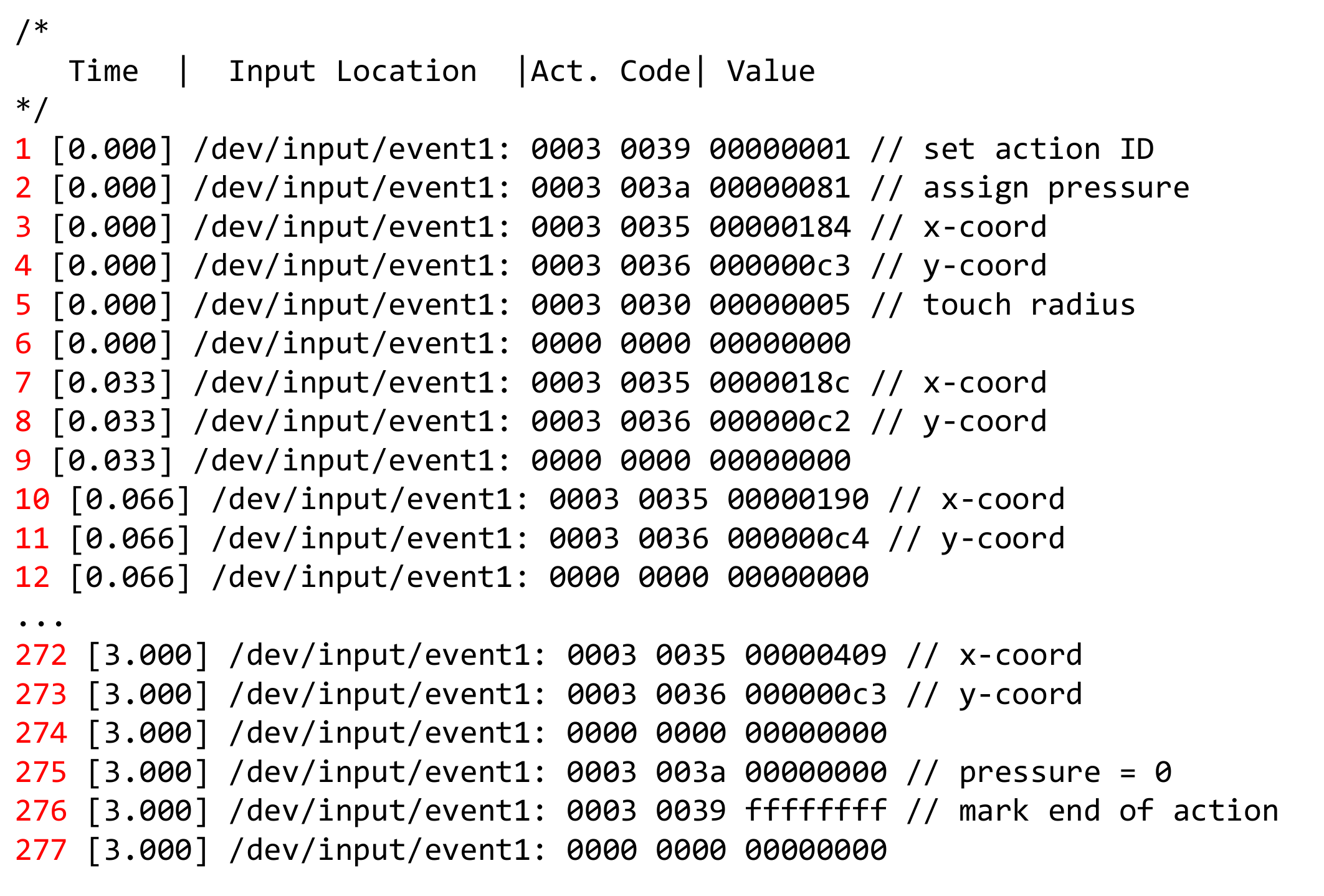}
        \caption{\texttt{One-Fingered Gesture} \texttt{sendevent} example}
        \label{fig:sfa-sendevent}
    \end{center}
\end{figure}

\noindent\textbf{Single-Fingered Actions.} To generate single-fingered events, each action is exported starting with the \texttt{start\_event} command (line 1 in Fig. \ref{fig:sfa-sendevent}). Then, for actions classified as a \texttt{Tap}, \vts provides a single $(x, y)$ coordinate pair, derived from the center of the detected bounding box of the \texttt{Tap}. For \texttt{Gesture}s, \vts iterates over each touch that makes up the \texttt{Gesture} action and appends the  $(x, y)$ pairs of each touch indicator to the list of instructions (lines 3-4, 7-8, \etc in Fig. \ref{fig:sfa-sendevent}). %
For \texttt{Long Tap}s, \vts performs similar processing to that of \texttt{Gesture}s, but instead uses only a single  $(x, y)$ pair from the initial detected touch indicator bounding box. %
Then, \vts ends the set of instructions for an action with the appropriate \texttt{end\_event} command (\eg lines 275 and 276 in Fig. \ref{fig:sfa-sendevent}).

\NEW{
\noindent\textbf{Multi-Fingered Actions.} %
\vts translates \textit{MFAs} differently compared to \textit{SFAs}. This is because, in order to achieve the intended replay behavior for a \textit{MFA}, all of the touches that occur within the same frame must be sent to and executed on the device at the same time. This means that, rather than translate one finger’s action in direct sequence as is accomplished for \textit{SFAs} (see Fig. \ref{fig:sfa-sendevent}), each finger’s action on the screen must be decomposed into its component touches per frame and translated along with the others that occur within the same frame. This requires including a command to alert the device which finger is being translated at each moment by indicating its slot value (\eg lines 7 and 14 in Fig. \ref{fig:mfa-sendevent})~\cite{multi-protocol}. 

Each finger’s action is exported beginning with the \texttt{start\_event} command (\eg lines 1 and 8 in Fig. \ref{fig:mfa-sendevent}). %
When one finger touch ends, \vts terminates this slot’s behavior by appending the \texttt{end\_event} command to the script (\eg lines 105-107 and 109-111 in Fig. \ref{fig:mfa-sendevent}). Because each finger's action is translated in association with a specific slot, separate finger’s actions are allowed to continue even if another finger's action ends. 
\begin{figure}[ht]
\begin{center}
		\includegraphics[width=\linewidth]{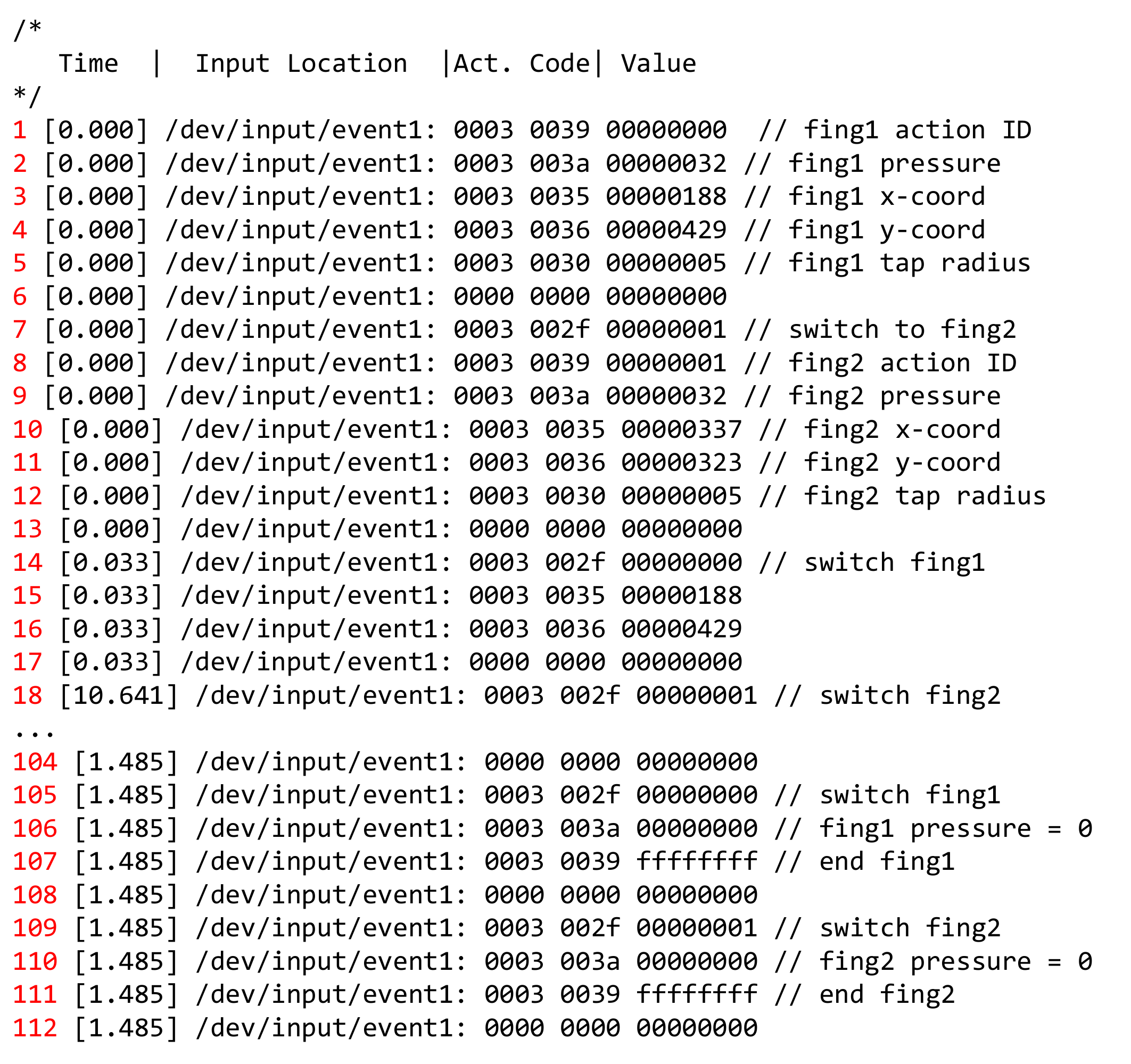}
        \caption{\texttt{Two-Fingered Gesture} \texttt{sendevent} example}
        \label{fig:mfa-sendevent}
    \end{center}
\end{figure}
}

\noindent For all \textit{SFAs} and \textit{MFAs}, the speed and duration of each instruction is extremely important in order to accurately replay the user's actions. To derive the speed and duration of these actions, \vts adds timestamps to each  $(x, y)$ touch location (see the \textit{Time} values in Fig. \ref{fig:sfa-sendevent} and Fig. \ref{fig:mfa-sendevent}) based on the timing between video frames (\ie for 30fps, there is a 33 millisecond delay between each frame), which will temporally separate each touch command sent to the device. %

And, in order to determine the delays between both \textit{SFAs} and \textit{MFAs}, the timing between video frames is again used. Our required 30fps frame-rate provides \vts with millisecond-level resolution of event timings. Higher frame-rates can increase the fidelity of replay timing.
 
\noindent\textbf{Scenario Replay.} Once all the actions have been converted into low-level \texttt{sendevent} instructions, they are written to a log file. This log file is then fed into a \textit{Translator} which converts the file into a runnable format that can be directly replayed on a device. This converted file along with a modified version of the RERAN engine~\cite{Gomez:ICSE13} is pushed to the target device. We optimized the original RERAN binary to replay event traces more efficiently. \NEW{Depending on the architecture of the target device, \vts provides a selection of RERAN binaries to allow for potential replay on a broader range of both physical and emulated devices\footnote{This has been particularly useful during the COVID-19 pandemic when physical access has been reduced.}}. Finally, the binary is executed using the converted file to faithfully replay the user actions originally recorded in the initial input video. We provide more examples of \vtss generated \texttt{sendevent} scripts, alongside our updated versions of the RERAN binaries in our online appendix~\cite{appendix}.

%% file: 4_implementation.tex
\NEW{\section{V2S+ Implementation}
\label{sec:implementation}
The original \ovts was implemented using a mix of Java, Python, and Bash, which made it difficult to use and manipulate to fit different research and development contexts. With this in mind, \vts was implemented entirely in Python with modularity in mind to encourage reuse and extension by future developers and researchers. %
Each phase and component can be easily substituted and/or extended by way of abstract classes which makes the entire pipeline malleable depending on the intended use case (\eg mobile application debugging, demonstration, \etc). In order to execute \vts, users specify the paths to the desired input video, \Faster, {\sc Opacity CNN}, \texttt{adb} binary, and the target device model in a modifiable \texttt{JSON} configuration file that is utilized throughout the pipeline. 
\subsection{Phase 1: Touch Detection}
This phase of \vts reads the path to the input video that depicts a natural application usage scenario from the configuration file. \vts defines and utilizes a distinct \textit{FrameExtractor (FE)} entity to prepare the video for touch detection. This class utilizes the \texttt{FFmpeg}~\cite{ffmpeg} tool to normalize the input video frame rate to 30fps and extract individual frames. Individual implementations of this portion of \vts can be substituted by implementing the \textit{AbstractVideoManipulator} class. 

Then, \vts allows its \textit{TouchDetector (TD)} class to utilize the pre-trained \Faster specified in the configuration file to locate individual touches within each of the frames extracted by the \textit{FE}. This \textit{TD} class can be reconfigured and customized by implementing the abstract \textit{AbstractTouchDetector} super class.

And finally, \vts utilizes an \textit{OpacityDetector (OD)} class, which mobilizes the {\sc Opacity CNN} specified in the configuration file, to classify each touch localized by the \textit{TD} class as having either a high or low opacity value. Again, this class's behavior can be modified or customized using the \textit{AbstractOpacityDetector} super class. 

The output of Phase 1 is a structured \texttt{JSON} file that contains all the detected touches details including their corresponding frame number, spatial location on the screen, and detected opacity.

\subsection{Phase 2: Action Classification}
This phase receives Phase 1's structured \texttt{JSON} output file as input. Then, \vtss \textit{GUIActionClassifier (GAC)} is responsible for detecting and classifying the \textit{SFAs} and \textit{MFAs} depicted in the input video based on the input from Phase 1. This can be altered by implementing the \textit{AbstractGUIActionClassifier} class. The output of Phase 2 is another structured \texttt{JSON} file that includes each classified action's component touches (or actions in the case of \textit{MFAs}). 

\subsection{Phase 3: Scenario Generation}
This phase utilizes Phase 2's \texttt{JSON} file as input. \vtss \textit{Action2EventConverter (A2EC)} then uses \texttt{sendevent} commands to translate the high-level actions detected by the \textit{GAC} into a low-level replayable output script. Depending on the specifications necessary to do so, developers can implement the \textit{AbstractAction2EventConverter} class to alter the behavior of this class. 

\vts then connects to the intended target device using the Android Debug Bridge (\texttt{adb}) and pushes the script to the device along with the appropriate pre-compiled RERAN~\cite{Gomez:ICSE13} binary. \vts then executes the replay and saves the recording as output to complete the pipeline.}

%% file: 5_study.tex
\vspace{-0.1cm}
\section{Empirical Evaluation}
\label{sec:study}
\NEW{In this section, we describe the methodology we used to evaluate \vts. The \textit{goal} of our empirical study is to assess the \textit{accuracy}, \textit{robustness}, \textit{performance}, and \textit{industrial utility} of \vts. %
The \textit{context} of this evaluation consists of: (i) sets of 15,000 and 10,000 images, corresponding to the evaluation of \vtss \Faster and {\sc Opacity CNN} respectively; (ii) a set of over \numApps Android applications including 64 of the top-rated apps from Google Play, 5 open source apps with real crashes, 5 open source apps with known bugs, 5 open source apps with controlled crashes, 14 hybrid apps, and 4 apps that have multi-touch capabilities; (iii) 2 popular physical target Android devices (Nexus 5 and Nexus 6P) and 1 emulated target Android device (Nexus~5). The main \textit{quality focus} of our study is the extent to which \vts can generate replayable scenarios that mimic original user app usages. %

To achieve our evaluation goals, we formulated the following six research questions:}

\begin{itemize}
	\item{\textit{\textbf{RQ$_1$}: How accurate is \approach in identifying the location of the touch indicator on the screen?}} %
	\item{\textit{\textbf{RQ$_2$}: How accurate is \approach in identifying the opacity of the touch indicator?}}
	\item{\textit{\NEW{\textbf{RQ$_3$}: How effective is \approach in generating a sequence of events that accurately mimics single-fingered user actions from video recordings of different applications?}}}
	\item{\textit{\NEW{\textbf{RQ$_4$}: How effective is \approach in generating a sequence of events that accurately mimics multi-fingered user actions from video recordings of different applications?}}}
	\item{\textit{\textbf{RQ$_5$}: What is \approach's overhead in terms of scenario generation?}}
	\item{\textit{\textbf{RQ$_6$}: Do practitioners perceive \vts as useful?}}
\end{itemize}

\subsection{RQ$_1$: Accuracy of Faster R-CNN}
\label{subsec:study-rq1}

To answer RQ$_1$, we first evaluated the ability of \vtss \Faster to accurately identify and localize the \textit{touch indicators} present in screen recording frames with \textit{bounding boxes}. To accomplish this, we followed the procedure to generate training data outlined in \secref{subsec:approach-detection} with the 70\%--30\% split for the training and testing sets, respectively. The implementation of the \Faster object detection used by \vts is coupled to the size of the images used for training and inference. Thus, to ensure \vtss model functions across different devices, we trained many \Faster models. 

For the original implementation of \vts, we trained two distinct \Faster models (\ie one for Nexus 5 and one for Nexus 6P). %
For all of these models, we resized the images from the \ReDraw dataset to the target device image size. As we show in the course of answering other RQs, we found that resizing the images in the already large \ReDraw dataset ~\cite{ReDrawD}%
, as opposed to re-collecting natively sized images for each device, resulted in highly accurate detections in practice. %

We used the \textit{TensorFlow Object Detection API}~\cite{TFODA} to train our model. %
Moreover, for the training process, we modified several of the hyper-parameters after conducting an initial set of experiments. These changes affected the number of classes (\ie 1) and the maximum number of detections per class or image (\ie 10). We also modified the learning rate for \vts after 50k (\ie $3\times10^{-5}$) and 100k (\ie $3\times10^{-6}$) iterations. %
Each training process was run for 150k steps with a batch size of 1, and our implementation of \Faster utilized a \VGGNet~\cite{Simonyan:ICLR14} instance pre-trained on the MSCOCO dataset~\cite{Lin:LNCS14}. We provide our full set of model parameters in our online appendix~\cite{appendix}.

To validate the accuracy of \vtss \Faster models we utilize \textit{Mean Average Precision} (mAP) which is commonly used to evaluate techniques for the object detection task. This metric is typically computed by averaging the \textit{precision} over all the categories in the data; however, given that we have a single class (the touch indicator icon), we report results only for this class. Thus, our mAP is computed as $mAP=TP/(TP+FP)$ where $TP$ corresponds to an identified image region with a correct corresponding label, and $FP$ corresponds to the identified image regions with the incorrect label, which in our case would be an image region that is falsely identified as a touch indicator. We use different values of the Intersection Over Union (IoU) \cite{Ren:NIPS15} between the region of the prediction and the ground truth region (IoU thresholds ranging from 0.5 to 0.95 with 0.05 increments), to determine if the prediction  is a TP or FP.
Additionally, we evaluate the \textit{Average Recall} of our model in order to determine if our model misses detecting any instances of the touch indicator. This is computed by $AR=TP/k$ where $TP$ is the same definition stated above, and $k$ corresponds to the total number of possible $TP$ predictions.

\begin{figure}[t]
    \centering
    \begin{subfigure}[b]{0.4\linewidth}
        \centering
        \includegraphics[height=0.3in]{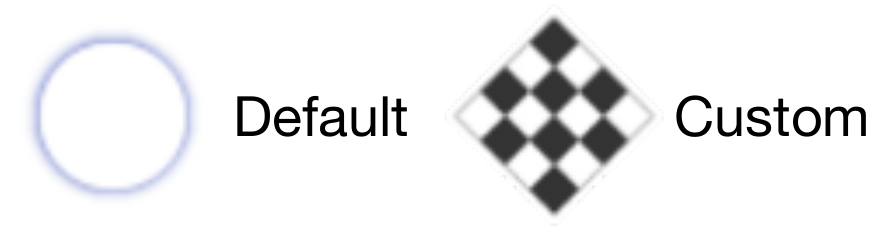}
        \caption{Touch indicators}
        \label{fig:touches}
        \vspace{-0.5em}
    \end{subfigure}%
     ~
    \begin{subfigure}[b]{0.5\linewidth}
        \centering
        \includegraphics[height=0.3in]{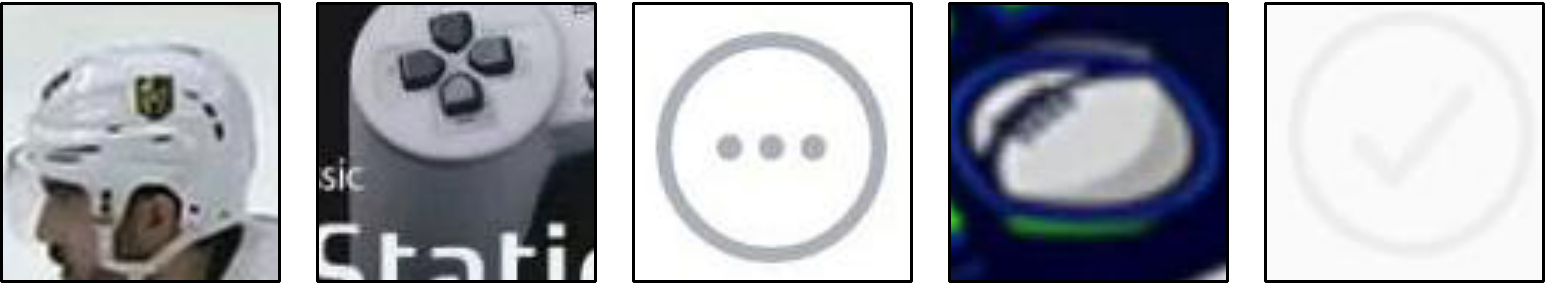}
        \caption{False positive detections}
        \label{fig:false_positives}
        \vspace{-0.5em}
    \end{subfigure}
    \caption{Touch indicators and failed detections}%
    \label{fig:touches_false_positives}
     \vspace{-0.5cm}
\end{figure}

During preliminary experiments with \Faster using the default \textit{touch indicator} (see \figref{fig:touches}), we found that, due to the default touch indicator's likeness to other icons and images present in apps, it was prone to very occasional false positive detections (\figref{fig:false_positives}). Thus, we analyzed particular cases in which the default touch indicator failed and replaced it with a more distinct, high-contrast touch indicator (see \figref{fig:touches}). We found that this custom touch indicator marginally improved the accuracy of our models. It should be noted that replacing the touch indicator on a device, requires the device to be rooted. While this is an acceptable option for most developers, it may prove difficult for end-users. However, even with the default touch indicator, \vtss \Faster model still achieves extremely high levels of accuracy. \NEWEST{For these reasons, we only evaluate the custom touch indicator for RQ$_1$ and RQ$_2$.}

\subsection{RQ$_2$: Accuracy of Opacity CNN} 
\label{subsection:study-rq2}

To answer RQ$_2$, we evaluated the ability of \vts's {\sc Opacity CNN} to predict whether the opacity of the touch indicator is solid or semi-transparent. To accomplish this, we followed dataset generation procedure outlined in \secref{subsec:approach-detection}, where  %
equal number of full and partial opacity examples are generated.
Thus, the generated dataset contains equal numbers of full and partial opacity examples for a total of 10k, which are evenly split into 70\%--30\% training and testing sets. We used the TensorFlow framework in combination with Keras to implement the {\sc Opacity CNN}. In contrast to the \Faster model  previously used, we do not need to create a separate model for each device. This is due to the \textit{touch indicator} being resized when fed into the {\sc Opacity CNN}. Similarly to the \Faster, we evaluate {\sc Opacity CNN} using \textit{mAP} across our two classes (solid and semi-transparent). %

\subsection{RQ$_3$: Replay Accuracy for Scenarios with Single-Fingered Actions}
\label{subsec:study-rq3}
\NEW{
To answer RQ$_3$, we carried out three different studies to assess the accuracy of \vts with \textit{SFAs} within different types of applications: 2 studies with native applications and 1 study with hybrid applications.} %

\subsubsection{Replay Accuracy on Native Applications} 

We carried out two studies designed to assess both the \textit{depth} and \textit{breadth} of \vtss abilities to reproduce user events depicted in screen recordings within native applications. The first, the \textit{Controlled Study}, measures the depth of \vtss abilities through a user study during which we collected real videos from end users depicting: bugs, real crashes, synthetically injected crashes, and normal usage scenarios for 20 apps (\eg Car Report, Another Monitor, \etc). %
Next in the \textit{Popular Applications Study} we measured the breadth of \vtss abilities by recording scenarios for a larger, more diverse set of 64 most popular apps from the Google Play (\eg Tasty, Twitter, Airbnb, \etc). %
We provide the full details of these apps in our online appendix~\cite{appendix}. %

\noindent\textbf{Controlled Study.} In this study, we considered four types of recorded usage scenarios depicting: (i) normal usages, (ii) bugs, (iii) real crashes, and (iv) controlled crashes. Normal usage scenarios refer to video recordings exercising different features on popular apps. Bug scenarios refer to video recordings that exhibit a bug on open source apps. Finally, controlled crashes refer to injected crashes into open source apps. This allows us to control the number of steps before the crash is triggered. 

For this study, eight participants including 1 undergraduate, 3 masters, and 4 doctoral students were recruited from William \& Mary (approved by the Protection of Human Subjects Committee (PHSC) at W\&M under protocol PHSC-2019-01-22-13374) to record the videos, with each participant recording eight separate videos, two from each of the categories listed above. Four participants recorded videos on the physical Nexus 5 and four used the physical Nexus 6P. This accounts for a total of 64 videos, from 20 apps evenly distributed across all scenarios. Before recording each app, participants were either asked to use the app to become familiar with it, or read a bug/crash report before reproducing the fault. All of the real bugs and crashes were taken from past studies on mobile testing and bug reporting~\cite{Moran:ICST16,Moran:FSE15,Chaparro:FSE'19}. \NEW{These videos depict a total of 1,141 \textit{SFAs}, with an average of $\approx 18$ \textit{SFAs} per video.} %

\noindent\textbf{Popular Applications Study.} In this study, we considered a larger and more diverse set of apps from Google Play. Specifically, we downloaded the two highest-rated apps from each of non-game categories (\ie 32) for a total of 64 applications (see Table \ref{tab:rq3-all-results}). 

Two of the authors then recorded two scenarios per app accounting for 32 apps each, one using a physical Nexus 5 and the other using a physical Nexus 6P. %
The authors strived to use the apps as naturally as possible, and this evaluation procedure is in line with past evaluations of Android record-and-replay techniques~\cite{Qin:ICSE16,Gomez:ICSE13}. The recorded scenarios represented specific use cases of the apps that exercise at least one of the major features, and were independent of one another. During our experiments, we noticed certain instances where our recorded scenarios were not replicable, either due to non-determinism or dynamic content (\eg random popups appearing). Thus, we discarded these instances and were left with 111 app usage scenarios from 60 apps. It is worth noting that it would be nearly impossible for existing techniques such as RERAN~\cite{Gomez:ICSE13} or Barista~\cite{Fazzini:ICST17} to reproduce the scenarios due to the nondeterminism of the dynamic content, hence our decision to exclude them. \NEW{These usage scenarios depict a total of 2,150 \textit{SFAs} and an average of $\approx 19$ \textit{SFAs} per usage video.}

\NEW{
\subsubsection{Replay Accuracy on Hybrid Applications} 
We carried out a study to assess \vtss ability to reproduce scenarios on hybrid applications, \ie those implementing features natively using Android's API and features using web-based technologies.  \NEW{This study is motivated by the increasing prevalence of hybrid applications in the Google Play store~\cite{Malavolta} and the current relative lack of GUI testing tools for these applications. 
We chose to evaluate \vtss performance with hybrid applications, separately from native applications, to demonstrate that \vts is one of few tools to support both hybrid and native apps. %
}

\noindent\textbf{Popular Applications Study.} In this study, we utilized 14 apps with hybrid features from 7 popular application categories -- 2 per category (see Table \ref{tab:rq3-169-old-model}). For each of these 14 apps, we devised 2 scenarios that interacted with the application's core hybrid functionality. %
We recruited 7 participants for the study (6 undergraduate students and 1 graduate student). Each of the 28 scenarios was recorded twice within the group, which allowed for each participant to record 8 different scenarios on a Nexus 5 emulator. We used an emulator (rather than a physical device) because of restrictions related to COVID-19. %
In an attempt to allow participants to interact with each application as naturally as possible, each participant was allowed to familiarize themselves with the application and with the scenario before recording. A total of 56 videos were collected for this study, but due to non-deterministic or dynamic content or \texttt{apk} issues, we discarded 3 videos. \NEW{These videos depict a total of 622 \textit{SFAs}, for an average of $\approx 11$ \textit{SFAs} per video.}

\subsubsection{Scenario Evaluation}\label{sec:scen_eval} \noindent 

\NEW{In addition to producing the video recordings, participants were asked to watch their scenarios and record their (ground truth) sequence of actions depicted in the videos. Participants were asked to classify each action depicted in each scenario as one of three options: (i) a \texttt{Tap}, an action where the touch indicator remains in approximately the same place and lasts for a short duration;  (ii) a \texttt{Long Tap}, an action where the touch indicator remains in about the same place for the entire duration of the action and the duration of the action lasts over a longer period than a \texttt{Tap} (\ie usually about 1 second or more); or  (iii) a \texttt{Gesture}, an action where the touch indicator moves about the screen that can be short or long in duration. %
An action sequence is, then, a sequence of action types (according to the three types defined above).
After the videos and action sequences were collected, two authors verified the correctness of:  (1) the recordings; and (2) the action sequences (compared to the recordings). 
Only in three cases,
we revised the action sequence to be properly aligned with the video and/or we trimmed out parts at the end or beginning of the video that were not relevant to app usage scenario, including cases where participants had included opening or closing the application or interacting with other applications in their recordings.}}%

To measure how accurately \vts replays videos that depict \textit{SFAs}, we use 3 different metrics. To compute these metrics, we manually derived the ground truth sequence of action types for each recorded video. First, we use Levenshtein distance, which is commonly used to compute distances between words at character level, to compare the original list of action types to the list of classified actions generated by \vts. Thus, we consider each type of action being represented as a character, and scenarios as sequences of characters which represent series of actions. A low Levenshtein distance value indicates fewer required changes to transform \vtss output to the ground truth set of actions. Additionally, we compute the longest common subsequence (LCS) to find the largest sequence of each scenario from \vtss output that aligns with the ground truth scenario from the video recording. For this LCS measure, the higher the percentage (between the LCS and the ground truth sequence), the closer \vtss trace is to a perfect match of the original trace. Moreover, we also computed the precision and recall for \vts to predict each type of action across all scenarios when compared to the ground truth. Finally, in order to validate the fidelity of the replayed scenarios generated by \vts compared to the original video recording, we manually compared each original video to each reproduced scenario from \vts, and determined the number of actions for each video that were faithfully replayed.

\NEW{\subsection{RQ$_4$: Replay Accuracy for Scenarios with Multi-Fingered Actions} To evaluate RQ$_4$, we conducted a study of popular applications involving one of the authors to demonstrate \vtss ability to correctly identify and replay \textit{MFAs} that are commonly seen in the wild. We studied 4 different native applications in 2 different categories in the Google Play store as shown in \tabref{tab:multi-apps}. %
These categories were chosen because they are known to host many applications that rely on \textit{MFAs} when using core app functionality. Within each of the 4 applications, we enumerated the supported \textit{MFAs} that result in specific behavior (also listed in \tabref{tab:multi-apps}) and designed scenarios that utilized these in both \textit{artificial} and \textit{natural  scenarios}. \textit{Artificial scenarios} are scenarios where the user simply executes multiple \textit{MFAs} in sequence; these types of scenarios were recorded to ensure that \vts can replicate the breadth of \textit{MFAs} that each application supports. \textit{Natural scenarios} are those that utilize basic functionality of the application and also incorporate one or more \textit{MFAs}; these were included in the study to ensure that \vts can accurately replay \textit{MFAs} when they occur in sequence with other actions supported by \vts (\eg \texttt{Tap}s). For the Art \& Design applications, we designed 1 natural scenario and 1 artificial scenario. For the Navigation applications, we designed 2 natural scenarios and 2 artificial scenarios. We chose to record more scenarios for Navigation apps than Art \& Design applications because there are more relevant multi-fingered actions in Navigation apps, as shown in Table \ref{tab:multi-apps}. %
\NEW{This study comprised 12 videos  that depicted 115 actions: 44 \textit{MFAs} and 71 \textit{SFAs}. These videos must include \textit{SFAs} in addition to MFAs because the natural scenarios required SFAs between the MFAs to create meaningful interactions with the applications. On average, each video contained $\approx 4$ \textit{MFAs} and $\approx 6$ \textit{SFAs}.}

One author of this paper recorded each of these 12 scenarios on a physical Nexus 5 Android device. This author familiarized herself with the applications and the scenarios before recording to ensure that the scenarios progressed in a natural fashion. Then, after filming the videos, this author rewatched each recording and classified each action either as a (i) \texttt{Tap}, %
a (ii) \texttt{Long Tap}, %
or a (iii) \texttt{Gesture}. %
These classifications were then used for empirically verifying the performance of \vts at replicating the sequences of gestures using the metrics listed in \secref{sec:scen_eval}.} %

\begin{table}[t]
\caption{Multi-Touch Study Applications}
\label{tab:multi-apps}
\centering
\footnotesize
\begin{tabular}{|l|l|l|}\hline
    \textbf{App Category} & \textbf{Apps} & \textbf{Actions Demonstrated}\\
    \hline
 & 
    \multirow{3}{*}{Infinite Painter} & Pinch to zoom in \\
               && Pull to zoom out \\
               && Two-finger rotation\\ \cline{2-3}& 
    \multirow{3}{*}{Sketchbook} & Pinch to zoom in \\
                && Pull to zoom out \\
    \multirow{-6}{*}{Art \& Design}  && Two-finger rotation\\ \hline
 & 
    \multirow{5}{*}{Google Maps} & Pinch to zoom in \\
                && Pull to zoom out \\
                &&Two-finger rotation\\
                && Two-finger tap\\
      && Two-finger "shove"\\
    \cline{2-3} & 
    \multirow{4}{*}{Waze} & Pinch to zoom in \\
              && Pull to zoom out \\
                && Two-finger rotation\\
    \multirow{-9}{*}{Navigation}            && Two-finger tap\\
     \hline
\end{tabular}
\end{table}

\subsection{RQ$_5$: Performance}
\label{subsec:study-rq4}

To address RQ$_5$, we evaluated \vts by measuring the average time it takes for a video to pass through each of the three phases of the \vts approach on commodity hardware (\ie a single NVIDIA GTX 1080Ti). We see this as a worst case scenario for \vts performance, as our approach could perform substantially faster on specialized hardware. Note that we modified the RERAN engine to run fast enough for more precise swipes. Therefore, we expect our \vts to have a lower overhead compare to the one reported in RERAN's respective paper \cite{Gomez:ICSE13}.

\subsection{RQ$_6$: Perceived Usefulness}

Ultimately, our goal is to integrate \vts into real-world development environments. Thus, as part of our evaluation, we investigated \vtss perceived usefulness with three developers who build Android apps for their respective companies.

The developers (\aka participants) were contacted through direct contact of the authors. Participant~\#1 (P1) was a front-end developer on the image search team of the Google Search app~\cite{g-image-search}, participant~\#2 (P2) is a developer of the 7-Eleven Android app~\cite{7-eleven}, and participant~\#3 (P3) is a backend developer for the Proximus shopping basket app~\cite{proximus}. We interviewed the participants using a set of questions organized in two sections. The first section aimed to collect information on participants' background, including their role at the company, the information used to complete their tasks, the quality of this information, the challenges of collecting it, and how they use videos in their every-day activities. The second section aimed to assess \vtss potential usefulness as well as its accuracy in generating replayable scenarios. This section also asked the participants for feedback to improve \vts including likert scale questions. We provide the complete list of interview questions used in our online appendix ~\cite{appendix}, and discuss selected questions in \secref{subsec:results-rq5}.

The participants answered questions from the second section by comparing two videos showing the same usage scenario for their respective app: one video displaying the scenario manually executed on the app, and the other one displaying the scenario executed automatically via \vtss generated script. Specifically, we defined, recorded, and manually executed a usage scenario on each app. Then, we ran \vts on the resulting video recordings. To define the scenarios, we identified a feature on each app involving any of the action types (\ie taps, long taps, and gestures). Then, we generated a video showing the original scenario (\ie video recording) and next to it the replayed scenario generated when executing \vtss script. Both recordings highlight the actions performed on the app. We presented the video to participants as well as \vtss script with the high-level actions automatically identified from the original video.

%% file: 6_results.tex
\section{Evaluation Results}
\label{sec:results}

We present and discuss the results of \vtss evaluation.

\subsection{RQ$_1$: Accuracy of \Faster}
\label{subsec:results-rq1}

\begin{table}[tb]
	\scriptsize
	\centering
	\caption{Touch Indicator Detection Accuracy - Original}
	\vspace{-0.9em}
	\label{tab:rq1_map_ar_og}
	\begin{tabular}{|l|c|c|c|c|} \hline
		\textbf{Model}&\textbf{Device}&\textbf{mAP}&\textbf{mAP@.75}&\textbf{AR}\\
		\hline
		\Faster-Original&Nexus 5&97.36\%&99.01\%&98.57\%\\
		\Faster-Original&Nexus 6P&96.94\%&99.01\%&98.19\%\\
		\Faster-Modified&Nexus 5&97.98\%&99.01\%&99.33\%\\
		\Faster-Modified&Nexus 6P&97.49\%&99.01\%&99.07\%\\
		\hline
	\end{tabular}
	\vspace{-0.3cm}
\end{table}

\tabref{tab:rq1_map_ar_og} depicts the precision and recall for \vtss \Faster network for touch indicator detection on different devices and datasets. The first column identifies the usage of either the default touch indicator or the modified version. The second column describes the target device for each trained model. \NEW{The third column provides the mAP, which is calculated based on the CoCo detection metric\footnote{https://cocodataset.org/\#detection-eval} that uses different values of the Intersection Over Union (IoU) \cite{Ren:NIPS15} between the area of the prediction and the ground truth area (IoU thresholds ranging from 0.5 to 0.95 with 0.05 increments). The fourth column shows the mAP when the IoU between the area of the prediction and the ground truth is above $75\%$. All models achieve a least $\approx$97\% mAP, indicating that $\approx$97\%  of the touch indicator predictions made by \vtss object detection network are correct.} mAP only improves when we consider bounding box IoUs that match the ground truth bounding boxes by at least 75\%, which illustrates that when the model is able to predict a reasonably accurate bounding box, it nearly always correctly detects the touch indicator (in $\approx$99\% of the cases). As illustrated by the last column in \tabref{tab:rq1_map_ar_og}, the model also achieves extremely high recall, detecting at least $\approx$98\% of the touch indicators. %

\vspace{0.2cm}
\begin{tcolorbox}[enhanced,skin=enhancedmiddle,borderline={1mm}{0mm}{MidnightBlue}]
\textbf{Answer to RQ$_1$}: \vts benefits from the strong performance of its object detection technique to detect touch indicators. All \Faster models achieved at least $\approx$ 97\% precision and at least $\approx$ 98\% recall across devices.%
\end{tcolorbox} 

\begin{table}[tb]
	\scriptsize
	\centering
	\caption{Confusion Matrix for Opacity CNN. Low Opacity Original (LO-Orig.), High Opacity Original (HO-Orig.), Low Opacity Custom (LO-Cust.), High Opacity Custom (HO-Cust.) }
	\vspace{-0.9em}
	\label{tab:rq2_cm_opacity}
	\begin{tabular}{|r|r|r|r|r|r|} \hline
		&\textbf{Total}&\textbf{LO-Orig.}& \textbf{HO-Orig.}&\textbf{LO-Cust.}&\textbf{HO-Cust.}\\
		\hline
		\textbf{Low Op}&5000&\cellcolor{MidnightBlue!20}97.8\%&2.2\%&\cellcolor{MidnightBlue!20}99.7\%&0.3\%\\
		\textbf{High Op}&5000&1.4\%&\cellcolor{MidnightBlue!20}98.6\%&0.8\%&\cellcolor{MidnightBlue!20}99.2\%\\
		\hline
	\end{tabular}
	\vspace{-0.3cm}
\end{table}

\subsection{RQ$_2$: Precision of the {\sc Opacity CNN}}
\label{subsec:results-rq2}

To illustrate the {\sc Opacity Network}'s precision in classifying the two opacity levels of touch indicators, we present the confusion matrix in \tabref{tab:rq2_cm_opacity}. The results are presented for both the default and modified (aka customized) touch indicator. The overall precision for the original touch indicator is 98.2\% whereas for the custom touch indicator is 99.4\%. These percentages are computed by dividing the correctly predicted label (i.e., Low/High-Opacity) by the number of members of that label for the original and custom touch indicators. Hence, it is clear \vtss~{\sc Opacity CNN} is highly precise at distinguishing between differing opacity levels.
\begin{tcolorbox}[enhanced,skin=enhancedmiddle,borderline={1mm}{0mm}{MidnightBlue}]
\textbf{Answer to RQ$_2$}: \vts benefits from the \Cnns precision in classifying levels of opacity. {\sc Opacity CNN} achieved an average precision above 98\% for both touch indicators.
\end{tcolorbox}  

\input{tab/eval.tex}

\vspace{-0.2cm}
\subsection{RQ$_3$: Replay Accuracy for Scenarios with Single-Fingered Actions}

We present and discuss the replay accuracy results of \vts  for scenarios with single-fingered actions, for both native and hybrid applications.

\label{subsec:results-rq3}
\subsubsection{Replay Accuracy on Native Applications}

\begin{figure}[t]
    \subfloat[LD-Study]{
    \begin{minipage}[c][1\width]{0.25\linewidth}
        \centering
        \includegraphics[height=0.8in]{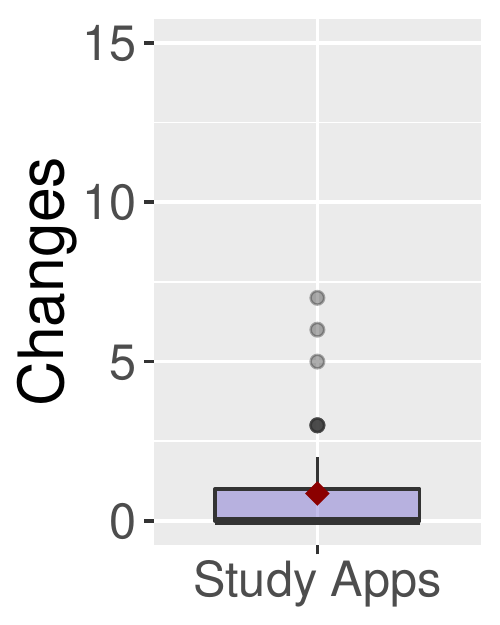}
        \label{fig:rq3_ld}
    \end{minipage}%
    }\subfloat[LD-Popular]{%
    \begin{minipage}[c][1\width]{0.25\linewidth}
        \centering
        \includegraphics[height=0.8in]{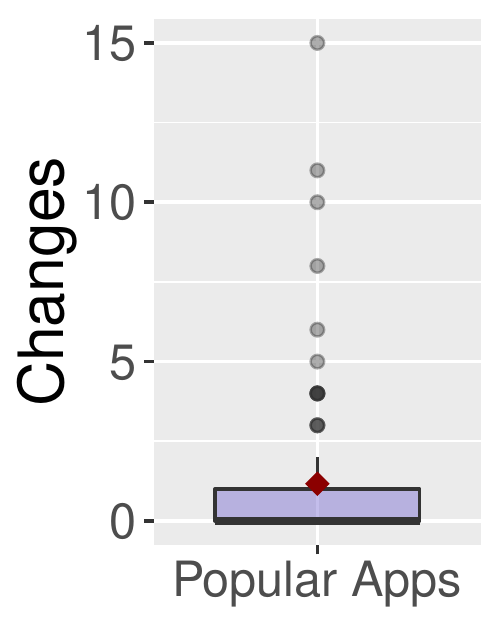}
        \label{fig:rq3_ld_p}
    \end{minipage}%
    }\subfloat[LCS-Study]{%
    
    \begin{minipage}[c][1\width]{0.25\linewidth}
        \centering
        \includegraphics[height=0.8in]{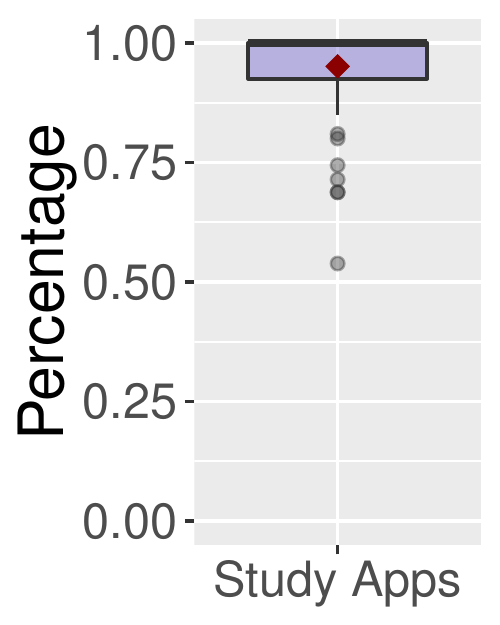}
        \label{fig:rq3_lcs}
    \end{minipage}
    }\subfloat[LCS-Popular]{%
    \begin{minipage}[c][1\width]{0.25\linewidth}
        \centering
        \includegraphics[height=0.8in]{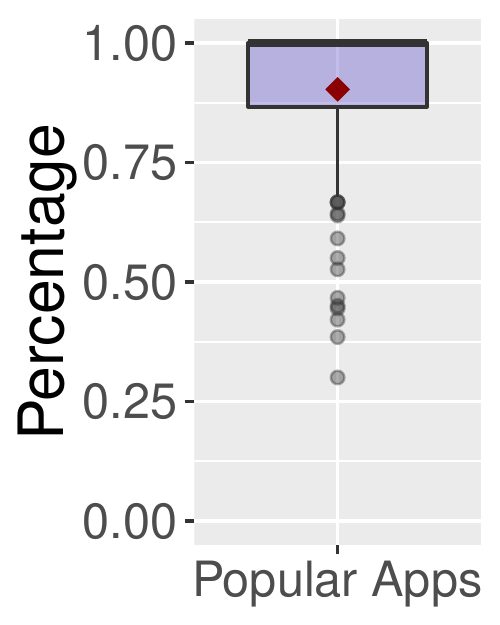}
        \label{fig:rq3_lcs_p}
    \end{minipage}
    }
    
    \caption{Effectiveness Metrics - Native Apps}%
    \label{fig:rq3_ld_lcs}
\end{figure}

\noindent \textbf{Levenshtein Distance.} \figref{fig:rq3_ld} and \ref{fig:rq3_ld_p} depict the number of changes required to transform the output event trace into the ground truth for the apps used in the \textit{Controlled Study} and the \textit{Popular Apps Study}, respectively. For the controlled study apps, on average it requires $0.85$ changes \NEW{(\ie \# of action changes)} per user trace to transform \vtss output into ground truth event trace, whereas for the popular apps it requires slightly more with $1.17$ changes. Overall, \vts requires minimal changes per event trace, being very similar to the ground truth. 

\noindent\textbf{Longest Common Subsequence.} %
\figref{fig:rq3_lcs} and \ref{fig:rq3_lcs_p} presents the percentage of events for each trace that match those in the original recording trace for the \textit{Controlled Study} and \textit{Popular Apps Study} respectively. On average \vts is able to correctly match 95.1\% of sequential events on the ground truth for the controlled study apps and 90.2\% for popular apps. These results suggest that \vts is able to generate sequences of actions that closely match the original trace in terms of action types. %

\begin{figure}[t]
    \begin{center}
		\includegraphics[width=\linewidth]{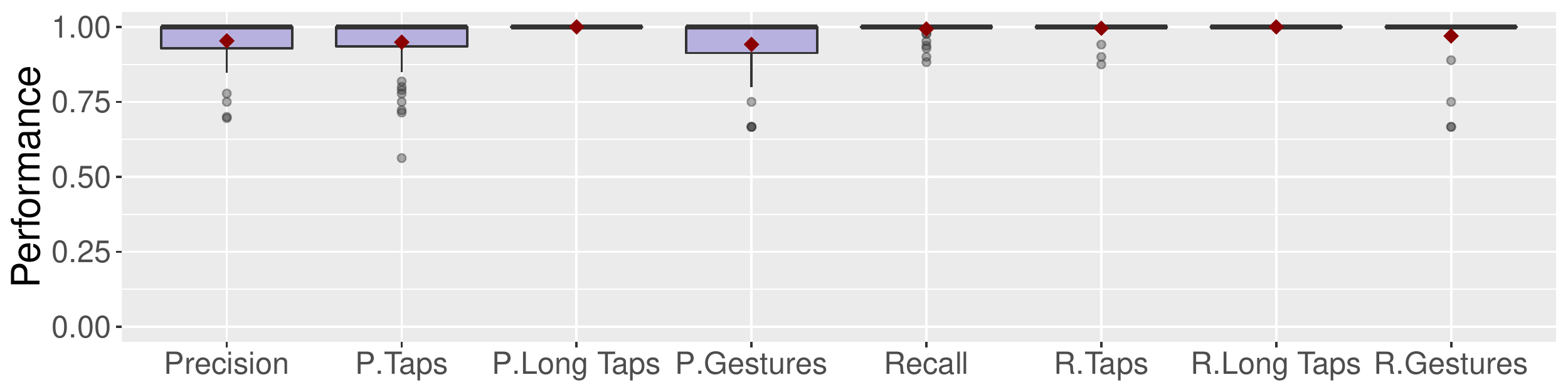}
		\vspace{-0.65cm}
        \caption{Precision and Recall - Controlled Study (Native Apps)}
        \label{fig:rq3_precision_recall}
    \end{center}
\end{figure}

\begin{figure}[t]
    \begin{center}
		\includegraphics[width=\linewidth]{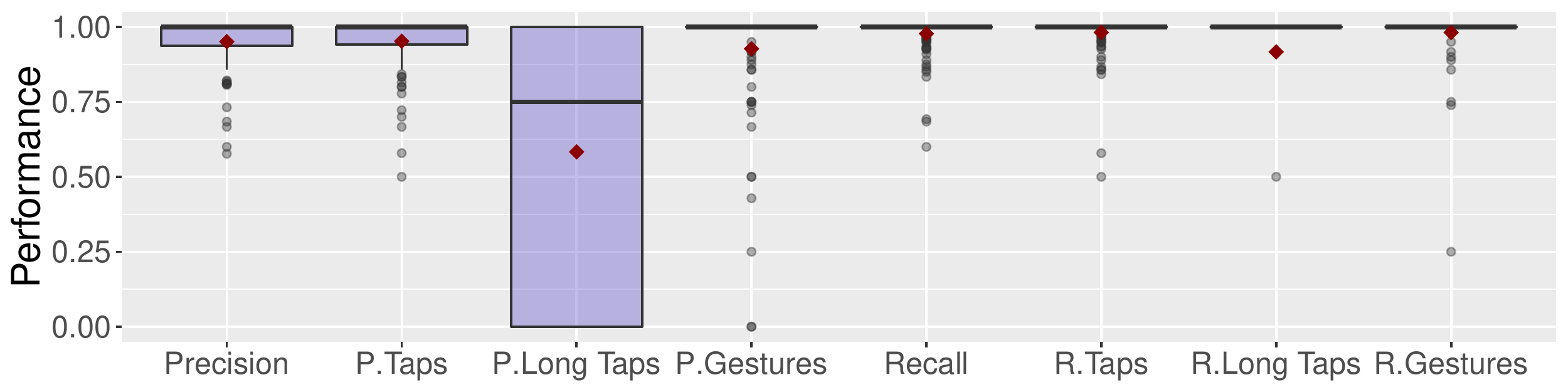}
		\vspace{-0.65cm}
        \caption{Precision and Recall - Popular Native Apps}
        \label{fig:rq3_precision_recall_p}
    \end{center}
\end{figure}

\noindent\textbf{Precision and Recall.} \figref{fig:rq3_precision_recall} and \ref{fig:rq3_precision_recall_p} show the precision and recall results for the \textit{Controlled Study} and \textit{Popular Apps Study}, respectively. These plots were constructed by creating an order agnostic ``bag of actions'' for each predicted action type, for each scenario in our datasets. Then, precision and recall are calculated by comparing the actions to a ground truth ``bag of actions'' and counting True/False Positives and False Negative. Finally, an overall average precision and recall are calculated across all action types. The results indicate that on average, the precision of the event traces is 95.3\% for the controlled study apps and 95\% for popular apps. This is also supported for each type of event showing also a high level of precision across types except for the precision on \texttt{Long Taps} for the popular apps. This is mainly due to the small number (\ie 9 \texttt{Long Taps}) of this event type across all the popular app traces. Also, \figref{fig:rq3_precision_recall}  and \ref{fig:rq3_precision_recall_p} illustrate that the recall across action types is high with an average of 99.3\% on controlled study apps and 97.8\% on the popular apps for all types of events. In general, we conclude that \vts can accurately predict the correct number of event types across traces.

\begin{figure}[t]
    \centering
    \begin{subfigure}[b]{0.25\linewidth}
        \centering
        \includegraphics[height=0.8in]{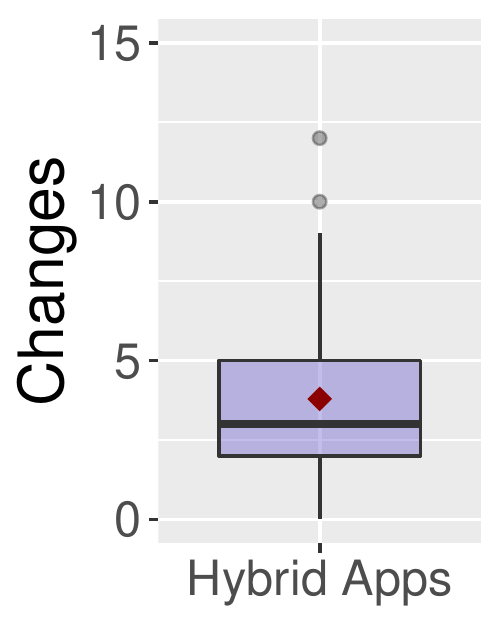}
        \caption{LD-Hybrid}
        \label{fig:rq3_ld_hybrid}
        \vspace{-0.5em}
    \end{subfigure}%
    \begin{subfigure}[b]{0.25\linewidth}
        \centering
        \includegraphics[height=0.8in]{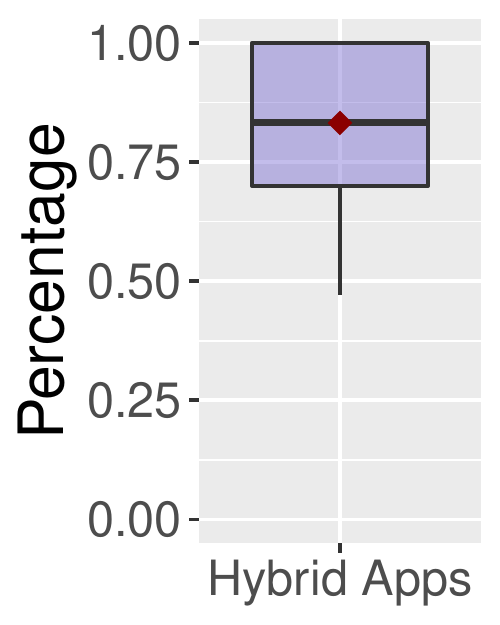}
        \caption{LCS-Hybrid}
        \label{fig:rq3_lcs_hybrid}
        \vspace{-0.5em}
    \end{subfigure}
    \vspace{1em}
    
    \caption{Effectiveness Metrics - Hybrid Apps}%
    \label{fig:rq3_ld_lcs_hybrid}
\vspace{-0.3cm}
\end{figure}

\input{tab/eval_169_old_model.tex}

\begin{figure}[t]
    \begin{center}
		\includegraphics[width=\linewidth]{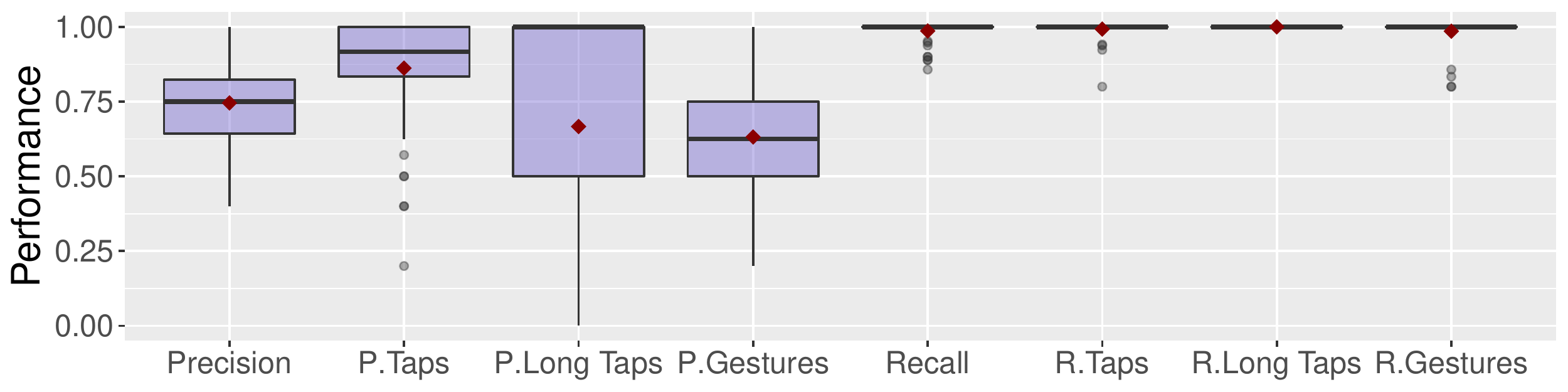}
		\vspace{-0.65cm}
        \caption{Precision and Recall - Hybrid Apps}
        \label{fig:rq3_precision_recall_hybrid}
    \end{center}
\vspace{-0.5cm}
\end{figure}

\noindent\textbf{Success Rate.} We also evaluated success rate of each replayed action for all scenarios across both RQ$_3$ studies. The 175 videos were analyzed manually and each action was marked as successful if the replayable scenario faithfully exercised the app features according to the original video. This means that in certain cases, videos will not \textit{exactly} match the original video recording (\eg due to a single keyboard keystroke error that still led to the same feature result). %

After validating all 64 videos for the controlled study, \vts fully reproduces 93.75\% of the scenarios, and 94.48\% of the consecutive actions. \vts fully reproduced 96.67\% of the scenarios for bugs and crashes and 91.18\% of apps usages. Detailed results for the \textit{Popular Apps Study} are shown in \tabref{tab:rq3-all-results}, where each app, scenario (with total number of actions), and successfully replayed actions are displayed. \textit{Green} cells indicate a fully reproduced video, \textit{Orange} cells indicate more than 50\% of events reproduced, and \textit{Red} cells indicate less than 50\% of reproduced events. \textit{Blue} cells show non-reproduced videos due to non-determinism/dynamic content. %
For the 111 scenarios recorded for the popular apps, \vts fully reproduced 81.98\% scenarios, and 89\% of the consecutive actions. Overall, this signals strong replay-ability performance across a highly diverse set of applications. Instances where \vts failed to reproduce scenarios are largely due to minor inaccuracies in \texttt{Gesture} events due to our video resolution of 30fps. We discuss potential solutions to this limitation in \secref{sec:limitations}.

\vspace{0.1cm}
\begin{tcolorbox}[enhanced,skin=enhancedmiddle,borderline={1mm}{0mm}{MidnightBlue}]
\textbf{Answer to RQ$_3$ (Native apps)}: \NEW{For native popular applications} used on physical devices, \vts is capable of generating event traces that require on average $\approx$ 1 change to match original user scenarios. Moreover,  at least 90.2\% of events match the ground truth, when considering the sequence of event types. Overall, precision and recall are $\approx$95\% and $\approx$98\% respectively for event types produced by \vts. Finally,  in 96.67\%  and 91.18\% of the cases, \vts successfully reproduces  bugs/crashes- and app-usage-related videos, respectively.
\end{tcolorbox}

\label{subsec:results-rq3_baseline}
\subsubsection{Replay Accuracy on Hybrid Applications}

\NEW{\noindent\textbf{Levenshtein Distance.} \figref{fig:rq3_ld_hybrid} depicts the number of changes required to transform \vtss event traces into the ground truth traces for apps used in the \textit{Hybrid Study}. In this study, 3.79 changes are required on average per user trace to transform \vtss output into the ground truth event trace. %

\noindent\textbf{Longest Common Subsequence.} %
\figref{fig:rq3_lcs_hybrid} presents the percentage of events for each trace that match those in the original recording trace for the \textit{Hybrid Study}. On average \vts is able to correctly match 83.19\% of sequential events on the ground truth for these hybrid applications. %

\noindent\textbf{Precision and Recall.} \figref{fig:rq3_precision_recall_hybrid} details the precision and recall results for the \textit{Hybrid Study}. 
The results indicate that on average, the precision of the event traces is 74.54\% for hybrid applications. \figref{fig:rq3_precision_recall_hybrid} also illustrates that the recall across action types is high with an average of 98.63\% for all types of events. 
}

\noindent\NEW{\textbf{Success Rate.} Finally, we also evaluated success rate of each replayed action in our \textit{Hybrid Study}. The 56 videos were analyzed manually and each action was marked as successful if the replayable scenario faithfully exercised the app features according to the original video. 

After validating all 56 videos for this study, 3 were removed because of non-determinism or problems with the \texttt{apks}. \vts fully reproduces 39 out of 53 scenarios (73.58\%), and 486 out of 622 of the consecutive actions (78.15\%). Detailed results are shown in \tabref{tab:rq3-169-old-model}, where each app, scenario (with total number of actions), and successfully replayed actions are displayed.} %

\noindent\NEW{\textbf{Discussion.} The results show that \vtss performance is lower for hybrid apps than for native apps. Due to restrictions related to COVID-19, we chose to conduct the hybrid-app study virtually and required participants to record scenarios on emulated devices on their personal machines. 
Our results analysis reveals that the use of these Android emulators in place of physical devices affected the quality of our results in two main ways: (i) utilizing a mouse or track pad to complete actions on an emulator is unnatural as it does not allow for the same ease of movement as a finger on a physical device, which resulted in more ``jagged" actions that are more often incorrectly classified and replayed by \vts; and (ii) the speed performance of the emulator is directly proportional to the speed of the computer's processing power, which manifested in our study as some participants' videos depicted ``laggy" actions that are not well-detected or replayed by \vts. We discuss potential solutions to these limitations in \secref{sec:limitations}.

\NEW{We
did not find any particular app-type-related factors that impacted the replay results for hybrid apps. As such, the gap in
our observed results (between hybrid and native apps) is likely not due to differences in the types of apps or \vtss ability to support them, but rather to
limitations of the video collection procedure.}}

\vspace{0.1cm}
\begin{tcolorbox}[enhanced,skin=enhancedmiddle,borderline={1mm}{0mm}{MidnightBlue}]
\NEW{\textbf{Answer to RQ$_3$ (Hybrid apps)}: For hybrid applications used on emulated devices, \vts is capable of generating event traces that require on average $\approx$ 3.5 changes to match original user scenarios. Moreover, at least 83\% of events match the ground truth when considering the sequence of event types. Overall, precision and recall are $\approx$ 75\% and $\approx$ 99\% respectively for event types produced by \vts. }%
\end{tcolorbox}

\NEW{\subsection{RQ$_4$: Replay Accuracy for Scenarios with Multi-Fingered Actions}}

We discuss the replay accuracy results of \vts  for scenarios with multi-fingered actions.

\begin{figure}[t]
    \centering
    \begin{subfigure}[b]{0.25\linewidth}
        \centering
        \includegraphics[height=0.8in]{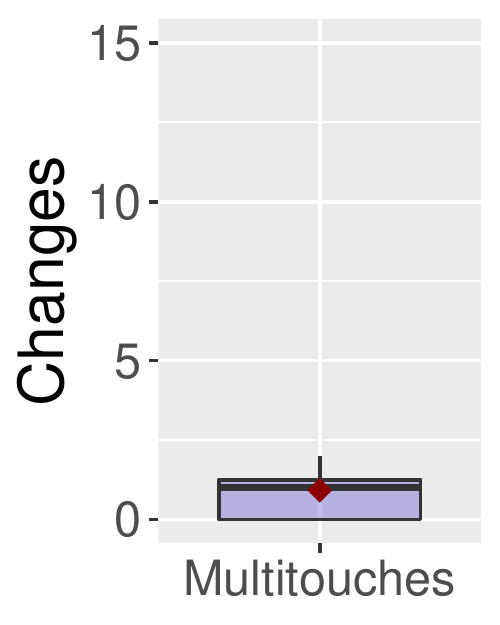}
        \caption{LD-Multi}
        \label{fig:rq3_ld_multi}
        \vspace{-0.5em}
    \end{subfigure}%
    \begin{subfigure}[b]{0.25\linewidth}
        \centering
        \includegraphics[height=0.8in]{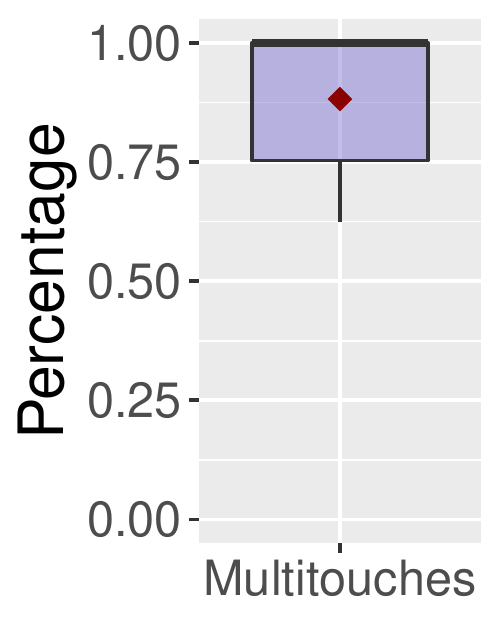}
        \caption{LCS-Multi}
        \label{fig:rq3_lcs_multi}
        \vspace{-0.5em}
    \end{subfigure}
    \vspace{1em}
    
    \caption{Effectiveness Metrics - Multi-Fingered Scenarios}%
    \label{fig:rq3_ld_lcs_multi}
\vspace{-0.3cm}
\end{figure}

\label{subsec:results-rq4}
\noindent\NEW{\textbf{Levenshtein Distance.} 
\figref{fig:rq3_ld_multi} depicts the number of changes required to transform the output event trace into the ground truth for the apps used in the \textit{Multi-Touch Study}. In this study, $\approx$ 0.92 changes are required on average per user trace to transform \vtss output into the ground truth event trace. %
If we separate the videos into their \textit{artificial} and \textit{natural} categories, \textit{artificial} scenarios only require on average $\approx$  0.33 changes per trace and \textit{natural} videos require on average $\approx$ 1.5 changes.
\NEW{If, instead, we separate the actions depicted in these videos into sequences of either \textit{SFAs} or \textit{MFAs}, then \textit{SFAs} require an average of one change per sequence and \textit{MFAs} require 0.083 changes per sequence, on average.}}

\noindent\NEW{\textbf{Longest Common Subsequence.} \figref{fig:rq3_lcs_multi} presents the percentage of sequential events for each trace that match those in the original recording trace for the \textit{Multi-Touch Study}. On average \vts is able to correctly match $\approx$ 88.2\% of sequential events on the ground truth for these multi-touch scenarios. If we separate the videos into their \textit{artificial} and \textit{natural} categories, \vts is able to correctly match $\approx$ 96.7\% of sequential events on the ground truth for \textit{artificial scenarios} and $\approx$ 79.7\% for \textit{natural scenarios}.

\NEW{If we separate \textit{SFAs} from \textit{MFAs} within these videos, \vts is able to correctly match $\approx$ 99.2\% of the sequential events in the ground truth for \textit{SFAs} and $\approx$ 97.9\% of the sequential events for \textit{MFAs}. According to our analysis, these results are higher than those obtained when aggregating these actions (reported above) for two main reasons.
First, considering \textit{MFAs} or \textit{SFAs} separately shortens the ground-truth action sequences, making the LCS metric more likely to be higher. Second, the videos included in this study were only able to depict \texttt{Two-Fingered Gestures} (as gestures involving more fingers do not produce meaningful application behavior), which also makes the \textit{MFA} sequences achieve a higher LCS.}

}

\begin{figure}[t]
    \begin{center}
		\includegraphics[width=\linewidth]{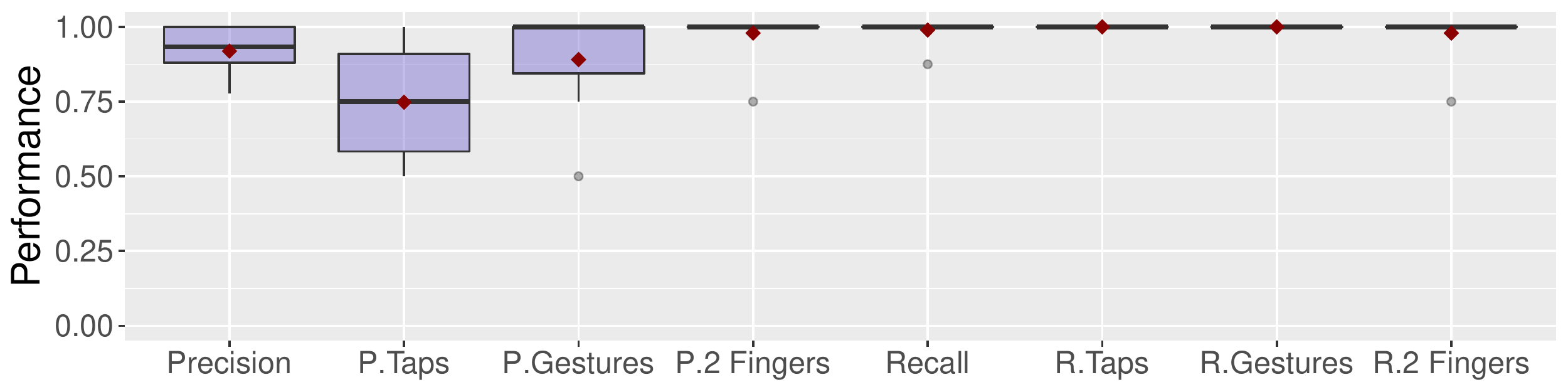}
		\vspace{-0.65cm}
        \caption{Precision and Recall - Multi-Fingered Actions}
        \label{fig:rq3_precision_recall_multi}
    \end{center}
\vspace{-0.5cm}
\end{figure}

\input{tab/eval_multi}

\noindent\NEW{\textbf{Precision and Recall.} \figref{fig:rq3_precision_recall_multi} details the precision and recall results for the \textit{Multi-Touch Study}. 
The results indicate that on average, the precision of the event traces is $\approx$ 91.9\% for multi-touch scenarios. \figref{fig:rq3_precision_recall_multi} also reveals that the recall across action types is very high with an average of $\approx$~99\% for all types of events in multi-touch scenarios. On average, \vts achieves $\approx$ 93.9\% precision and 100\% recall for \textit{artificial scenarios}, and the $\approx$ 89.9\% precision and $\approx$ 97.9\% recall for \textit{natural scenarios}. \NEWR{If we separate the actions depicted into these videos for \textit{SFAs} and \textit{MFAs}, we find that, for \textit{SFAs}, precision is $\approx 83\%$ and recall is $100\%$. For \textit{MFAs}, precision is $\approx 97.9\%$ and recall is $\approx 97.9\%$. }
}

\noindent\NEW{\textbf{Success Rate.} Finally, we also evaluated success rate of each replayed action in our \textit{Multi-Touch Study}. 
After manually validating the 12 videos for this study, we found that, on average, \vts fully reproduces 83.3\% of the scenarios, and 94.8\% of the consecutive actions. For \textit{artificial scenarios}, \vts is able to reproduce 83.3\% of the scenarios and 95.7\% of the consecutive actions. For \textit{natural scenarios}, \vts is able to reproduce 83.3\% of the scenarios and 94.5\% of sequential actions. %
Detailed results are shown in \tabref{tab:rq3-all-results-multi}, where each app, scenario (with total number of actions), and successfully replayed actions are displayed.} \NEWR{Because this metric is based on the progression of the depicted scenario as a whole, we were unable to separate \textit{SFAs} from \textit{MFAs} to compute the metrics.}

\vspace{0.1cm}
\begin{tcolorbox}[enhanced,skin=enhancedmiddle,borderline={1mm}{0mm}{MidnightBlue}]
\NEW{\textbf{Answer to RQ$_4$}: \vts is capable of generating event traces involving multi-fingered actions that require on average $<$ 1 change to match original user scenarios. Moreover,  at least 88\% of events match the ground truth, when considering the sequence of event types. Overall, precision and recall are $\approx$ 92\% and $\approx$ 99\% respectively for all event types reproduced by \vts.
 \vts successfully  reproduces $\approx$ 83.3\% of multi-fingered scenarios.}
\end{tcolorbox}

\subsection{RQ$_5$: Approach Performance}
\label{subsec:results-rq5}

To measure the performance of \vts, we measured the average time in seconds/frame (s/f) for a single video frame to be processed across all recorded videos for three components: (i) the frame extraction process (0.045 s/f), (ii) the touch indicator detection process (1.09 s/f), and (iii) the opacity classification process (0.032 s/f). The script generation time is negligible compared to these other processing times, and is only performed once per video. This means that an average video around 3 mins in length\footnote{This is a duration longer than most of the scenarios recorded for our evaluation.} would take \vts $\approx$105 minutes to fully process and generate the script. %
However, this process is \textit{fully automated}, can run in the background, and can be accelerated by more advanced hardware. We expect the overhead of our replayed scripts to be similar or better than RERAN since \vts replay engine is essentially an improved version of RERAN's. %

\begin{tcolorbox}[enhanced,skin=enhancedmiddle,borderline={1mm}{0mm}{MidnightBlue}]
\textbf{Answer to RQ$_5$}: \vts is capable of fully processing an average 3-min screen recording in  $\approx$105 mins. %
\end{tcolorbox} 
\subsection{RQ$_6$: Perceived Usefulness}
\label{subsec:results-rq6}

We discuss the results of the user study we conducted to assess \vtss perceived accuracy and usefulness.

First of all, the three industry participants (strongly) agreed that the scenarios produced by \vts (in the generated scripts) are accurate with respect to the scenarios that were manually executed.

The participants also agreed that further tool support is needed for helping QA members and other stakeholders with generating video recordings. For example,  P3 mentions that  while videos are "more useful than images" (in some cases), they ``may be difficult to record'' because of ``time constraints''. P1 remarked that the QA team could use \vts to help them create videos more optimally. P2 supported this claim as he mentions that \vts could help ``the QA team write/provide commands or steps, then the tool would read and execute these while recording a video of the scenario and problem. This solution could be integrated in the continuous integration pipeline''. In addition, P3 mentions that \vts could be used during app demos: \vts could ``automatically execute a script that shows the app functionality and record a video. In this way, the demo would focus on business explanation rather than on manually providing input to the app or execute a user scenario''.

P2 also indicated that \vts could be used to analyze user behavior within the app, which can help improve certain app screens and navigation. He mentions that \vts ``could collect the type of interactions, \# of taps, \etc to detect, for example, if certain screens are frequently used or if users often go back after they navigate to a particular screen''. He mentions that this data ``could be useful for marketing purposes''. P3 finds \vts potentially useful for helping reproduce hard-to-replicate bugs.

The participants provided valuable and specific feedback for improving \vts. They suggested to enrich the videos produced when executing \vtss script with a bounding box of the GUI components or screen areas being interacted with at each step. They also mention that the video could show (popup) comments that explain what is going on in the app (\eg a comment such as ``after 10 seconds, button X throws an error''), which can help replicate bugs. They would like to see additional information in the script, such as GUI metadata that provides more in-depth and easy-to-read information about each step. For example, the script could use the names or IDs of the GUI components being interacted with and produce steps such as ``the user tapped on the send button'' instead of ``the user tapped at (10.111,34.56)''. P3 mentioned that ``it would be nice to change the script programmatically by using the GUI components' metadata instead of coordinates, so the script is easier to maintain''. They suggest to include an interpreter of commands/steps, written in a simple and easy-to-write language, which would be translated into low-level commands.

\begin{tcolorbox}[enhanced,skin=enhancedmiddle,borderline={1mm}{0mm}{MidnightBlue}]
	\textbf{Answer to RQ$_6$}: Developers find \vts accurate in replicating app usage scenarios from input videos, and potentially useful for supporting several development tasks, including automatically replicating bugs, analyzing usage app behavior, helping users/QA members generate high-quality videos, and automating scenario executions.
\end{tcolorbox}

%% file: tab/eval.tex
\begin{table*}[t]
\fontsize{6}{7}\selectfont
\centering
\caption{Detailed Results for RQ$_3$ popular applications study. Green cells indicate fully reproduced videos, Orange cells $>$50\% reproduced, and Red Cells$ <$50\% reproduced. Blue cells show non-reproduced videos due to non-determinism/dynamic content.}
\vspace{-.8em}
\label{tab:rq3-all-results}
\begin{tabular}{|l|
>{\columncolor[HTML]{9AFF99}}l |
>{\columncolor[HTML]{9AFF99}}l |l|
>{\columncolor[HTML]{9AFF99}}l |l|l|
>{\columncolor[HTML]{9AFF99}}l |
>{\columncolor[HTML]{9AFF99}}l |l|
>{\columncolor[HTML]{9AFF99}}l |
>{\columncolor[HTML]{9AFF99}}l |l|
>{\columncolor[HTML]{9AFF99}}l |
>{\columncolor[HTML]{9AFF99}}l |}
\hline
\textbf{AppName} & \multicolumn{2}{l|}{\textbf{Rep. Actions}} & \textbf{AppName} & \multicolumn{2}{l|}{\textbf{Rep. Actions}} & \textbf{App Name} & \multicolumn{2}{l|}{\textbf{Rep. Actions}} & \textbf{App Name} & \multicolumn{2}{l|}{\textbf{Rep. Actions}} & \textbf{App Name} & \multicolumn{2}{l|}{\textbf{Rep. Actions}} \\ \hline
Ibis Paint X                    & 11/11                           & 36/36                             & Firefox                         & 22/22                            & \cellcolor[HTML]{D9E0F2}N/A      & Tasty                            & 20/20                            & \cellcolor[HTML]{FFCB2F}14/36    & SoundCloud                       & 12/12                            & 13/13                            & LetGo                            & 17/17                             & 15/15                           \\ \hline
Pixel Art Pro                   & 20/20                           & 9/9                               & MarcoPolo                       & 12/12                            & \cellcolor[HTML]{9AFF99}30/30    & Postmates                        & 26/26                            & 12/12                            & Shazam                           & \cellcolor[HTML]{FFCB2F}12/15    & 20/20                            & TikTok                           & 14/14                             & 11/11                           \\ \hline
Car-Part.com                    & 20/20                           & 16/16                             & Dig                             & 13/13                            & \cellcolor[HTML]{D9E0F2}N/A      & Calm                             & 9/9                              & \cellcolor[HTML]{FFCB2F}11/16    & Twitter                          & 14/14                            & 19/19                            & LinkedIn                         & 18/18                             & 13/13                           \\ \hline
CDL Practice                    & 8/8                             & 13/13                             & Clover                          & 15/15                            & \cellcolor[HTML]{9AFF99}19/19    & Lose It!                         & 36/36                            & \cellcolor[HTML]{D9E0F2}N/A      & News Break                       & \cellcolor[HTML]{F56B00}1/18     & 9/9                              & CBSSports                        & 25/25                             & 16/16                           \\ \hline
Sephora                         & \cellcolor[HTML]{FFCC67}4/9     & 13/13                             & PlantSnap                       & 39/39                            & \cellcolor[HTML]{FFCB2F}18/24    & U Remote                 & 14/14                            & 18/18                            & FamAlbum                      & 19/19                            & \cellcolor[HTML]{FFCB2F}8/26     & MLBatBat                         & \cellcolor[HTML]{FFCB2F}11/13     & \cellcolor[HTML]{D9E0F2}N/A     \\ \hline
SceneLook                       & 14/14                           & 16/16                             & Translator                      & 20/20                            & \cellcolor[HTML]{9AFF99}28/28    & LEGO                       & 52/52                            & 24/24                            & Baby-Track                     & 14/14                            & 12/12                            & G-Translate                 & \cellcolor[HTML]{FFCB2F}14/17     & 15/15                           \\ \hline
KJ Bible                        & 16/16                           & 19/19                             & Tubi                            & \cellcolor[HTML]{FD6864}2/19     & \cellcolor[HTML]{9AFF99}30/30    & Dev Libs                         & 35/35                            & 22/22                            & Walli                            & 8/8                              & \cellcolor[HTML]{D9E0F2}N/A      & G-Podcast                  & 9/9                               & 15/15                           \\ \hline
Bible App                       & 12/12                           & 15/15                             & Scan Radio                   & \cellcolor[HTML]{FFCC67}24/27    & \cellcolor[HTML]{D9E0F2}N/A      & Horoscope                        & 24/24                            & 19/19                            & ZEDGE                            & 9/9                              & 18/18                            & Airbnb                           & 9/9                               & 14/14                           \\ \hline
Indeed Jobs                     & 15/15                           & 19/19                             & Tktmaster                    & 30/30                            & \cellcolor[HTML]{9AFF99}14/14    & Waze                             & 17/17                            & 19/19                            & G-Photo                    & 18/18                            & 18/18                            & G-Earth                     & \cellcolor[HTML]{F56B00}8/13      & \cellcolor[HTML]{F56B00}1/30    \\ \hline
UPS Mobile                      & 16/16                           & \cellcolor[HTML]{FFCB2F}19/24     & Greet Cards                   & 23/23                            & \cellcolor[HTML]{D9E0F2}N/A      & Transit                          & 26/26                            & 18/18                            & PicsArt                          & 18/18                            & 39/39                            & DU Record                      & 15/15                             & 9/9                             \\ \hline
Webtoon                         & 17/17                           & \cellcolor[HTML]{FFCB2F}15/21     & QuickBooks                      & 47/47                            & \cellcolor[HTML]{9AFF99}28/28    & WebMD                            & \cellcolor[HTML]{FD6864}7/34     & \cellcolor[HTML]{FD6864}7/26     & G-Docs                      & \cellcolor[HTML]{F56B00}3/26     & \cellcolor[HTML]{D9E0F2}N/A      & AccuWeath                      & 13/13                             & 21/21                           \\ \hline
MangaToon                       & 16/16                           & 28/28                             & Yahoo Fin                   & 23/23                            & \cellcolor[HTML]{D9E0F2}N/A      & K-Health                         & 10/10                            & \cellcolor[HTML]{FFCB2F}15/24    & M. Outlook                     & 27/27                            & \cellcolor[HTML]{FFCB2F}21/26    & W. Radar                    & 14/14                             & 13/13                           \\ \hline
\end{tabular}
\end{table*}

%% file: tab/eval_169_old_model.tex
\begin{table}[b]
\fontsize{6}{7}\selectfont
\centering
\caption{Detailed Results for RQ$_3$ hybrid scenarios study. Green cells indicate fully reproduced videos, Orange cells $>$50\% reproduced, and Red Cells$ <$50\% reproduced. Blue cells show non-reproduced videos due to non-determinism/dynamic content or \texttt{apk} issues.}
\vspace{-.8em}
\label{tab:rq3-169-old-model}
\begin{tabular}{|l|
>{\columncolor[HTML]{9AFF99}}l |
>{\columncolor[HTML]{9AFF99}}l |
>{\columncolor[HTML]{9AFF99}}l|
>{\columncolor[HTML]{9AFF99}}l |}
\hline
% \cellcolor[HTML]{FD6864} red
% \cellcolor[HTML]{FFCC67} orange
% \cellcolor[HTML]{D9E0F2} grey
\textbf{AppName} & \multicolumn{2}{l|}{\textbf{Scen. 1 Rep. Actions}} &  \multicolumn{2}{l|}{\textbf{Scen. 2 Rep. Actions}}\\ \hline

Wikihow & 4/4& \cellcolor[HTML]{FD6864} 1/7&
\cellcolor[HTML]{FD6864} 3/9& 10/10\\ \hline

Wikipedia & \cellcolor[HTML]{FD6864}  3/14 & 13/13&
23/23& 17/17\\ \hline

Platt  & \cellcolor[HTML]{FD6864} 2/10& 14/14&
3/3& 3/3\\ \hline

USPS & 16/16 & 16/16&
24/24& \cellcolor[HTML]{FFCC67}18/23\\ \hline

CNBC & 10/10& 10/10&
11/11& 14/14\\ \hline

Seeking Alpha& 7/7& 5/5&
\cellcolor[HTML]{FD6864}0/5& \cellcolor[HTML]{FD6864}1/7\\ \hline

Old Navy & \cellcolor[HTML]{FFCC67}7/9& \cellcolor[HTML]{FD6864}1/11 &
\cellcolor[HTML]{FD6864}0/6& \cellcolor[HTML]{D9E0F2}N/A\\ \hline

Pinterest & 22/22& \cellcolor[HTML]{FD6864}0/21&
24/24& 24/24\\ \hline

Epocrates & \cellcolor[HTML]{FFCC67}10/13 &  \cellcolor[HTML]{FFCC67}12/17&
10/10& 6/6\\ \hline

Leafly & 18/18 &11/11&
 \cellcolor[HTML]{D9E0F2}N/A& 9/9\\ \hline

Newsbreak  & 5/5& 5/5&
9/9& 5/5\\ \hline

Guardian  &  \cellcolor[HTML]{FFCC67}5/9&  \cellcolor[HTML]{FD6864}1/8&
9/9& 9/9\\ \hline

Rainbow & 12/12&  \cellcolor[HTML]{FD6864}2/19&
12/12& 11/11\\ \hline

VS & 12/12&  \cellcolor[HTML]{FD6864}5/11&
 \cellcolor[HTML]{FD6864}2/10&  \cellcolor[HTML]{D9E0F2}N/A\\ \hline

\end{tabular}
\end{table}

%% file: tab/eval_multi.tex
\begin{table}[t]
\fontsize{6}{7}\selectfont
\centering
\caption{Detailed Results for RQ$_4$ multi-touch scenarios study. Green cells indicate fully reproduced videos, Orange cells $>$50\% reproduced, and Red Cells$ <$50\% reproduced. Blue cells show non-reproduced videos due to non-determinism/dynamic content or \texttt{apk} issues.}
\vspace{-.8em}
\label{tab:rq3-all-results-multi}
\begin{tabular}{|l|
>{\columncolor[HTML]{9AFF99}}l |
>{\columncolor[HTML]{9AFF99}}l |}
\hline
% \cellcolor[HTML]{FD6864} red
% \cellcolor[HTML]{FFCC67} orange
% \cellcolor[HTML]{D9E0F2} grey
\textbf{AppName} & \cellcolor[HTML]{FFFFFF}\textbf{Art. Scen.}& \cellcolor[HTML]{FFFFFF}\textbf{Nat. Scen.}\\ \hline

Infinite Painter & 5/5&19/19 \\\hline

Sketchbook & 5/5& 19/19\\\hline
    
              &\cellcolor[HTML]{FFCC67}4/5&15/15 \\\cline{2-3}
    \multirow{-2}{*}{Google Maps}&4/4& \cellcolor[HTML]{FD6864}3/8\\\hline
    
             &3/3&13/13\\\cline{2-3}
    \multirow{-2}{*}{Waze} &2/2&17/17\\\hline

\end{tabular}
\end{table}

%% file: 7_related-work.tex
\vspace{-0.2cm}
\section{Related Work}
\label{sec:related-work}
\vspace{0.1cm}

We briefly discuss prior research work related to \vts. %

\subsection{Analysis of video and screen captures}
Lin \etal \cite{Lin:NDSS14} proposed an approach called Screenmilker to automatically extract screenshots of sensitive information (\eg user entering a password) by using the Android Debug Bridge. %
This technique focuses on the extraction of keyboard inputs from "real-time" screenshots. Screenmilker is primarily focused upon extracting sensitive information, whereas \vts analyzes every single frame of a video to generate a high fidelity replay script from a sequence of video frames.

Krieter \etal \cite{Krieter:MobileHCI18} use video analysis to extract high-level descriptions of events from user video recordings on Android apps. Their approach generates log files that describe what events are happening at the app level. %
Compared to our work, this technique is not able to produce a script that would automatically replay the actions on a device, but instead simply describe high-level app events (\eg \textit{``WhatsApp chat list closed''}). Moreover, our work focuses on video analysis to help with bug reproduction and generation of test scenarios, rather than describing usage scenarios at a high level.

Bao \etal \cite{Bao:ICSE15} and Frisson \etal \cite{Frisson:CHI16} focus on the extraction of user interactions to facilitate behavioral analysis of developers during programming tasks using CV techniques. In our work, rather than focusing upon recording developers interactions, we instead focus on understanding and extracting generic user actions on mobile apps in order to generate high-fidelity replay scripts.

Other researchers have proposed approaches that focus on the generation of source code for Android applications from screenshots or mock-ups. These approaches rely on techniques that vary solely from CV-based \cite{Nguyen:ASE15} to DL-based \cite{Beltramelli:EICS18,Moran:TSE18,Chen:ICSE18}.

The most related work to \vts is the AppFlow approach introduced by Hu \etal \cite{Hu:FSE18}. AppFlow leverages machine learning techniques to analyze Android screens and categorize types of test cases that could be performed on them (\ie a sign in screen whose test case would be a user attempting to sign in). However, this technique is focused on the generation of semantically meaningful test cases in conjunction with automated dynamic analysis. In contrast, \vts is focused upon the automated replication of any type of user interaction on an Android device, whether this depicts a usage scenario or bug. Thus, \vts could be applied to automatically reproduce crowdsourced mobile app videos, whereas AppFlow is primarily concerned with the generation of tests rather than the reproduction of existing scenarios.

Other work has focused on detecting duplicate video-based bug reports. For example, Cooper \etal ~\cite{tango} proposed an approach that processes each new bug report that is submitted and automatically creates a list of the top 5 most similar entries, marking the report in the issue tracker as a likely duplicate. Similar to \vts, TANGO utilizes popular CV techniques to analyze the video inputs. However, TANGO also relies on the textual element of submitted bug reports to draw conclusions in a way that \vts does not, using text retrieval and character recognition tools to analyze this additional information. Kordopatis-Zilos \etal ~\cite{kz-cnn} introduced an approach that utilizes CNNs to extract near-identical videos that could be included in bug reports. This framework makes use of three distinct deep networks and two feature aggregation techniques to extract and make sense of visual features and collect  metadata that can then be used to compare videos. Much like \vts, this approach utilizes CNN frameworks to turn visual information into data that can be processed and composed in the context of the problem at hand. However, neither of the above approaches accomplish the same action classification or replay techniques implemented in \vts.  %

\subsection{Record and replay}
Many tools assist in recording and replaying tests for mobile platforms \cite{Gomez:ICSE13, airtest, replaykit, Jeon:12,Moran:MOBILESoft'17}. However, many of these tools require the recording of low-level events using \texttt{adb}, which usually requires rooting of a device, or loading a custom operating system (OS) to capture user actions/events that are otherwise not available through standard tools such as \texttt{adb}.
While our approach uses RERAN \cite{Gomez:ICSE13} to replay system-level events, we rely on video frames to transform touch overlays to low-level events. This facilitates bug reporting for users by minimizing the requirement of specialized programs to record and replay user scenarios.%

Hu \etal \cite{Hu:OOPSLA15} developed VALERA for replaying device actions, sensor and network inputs (\eg GPS, accelerometer, etc.), event schedules, and inter-app communication. This approach requires a rooted target device and the installation of a modified Android runtime environment. These requirements may lead to practical limitations, such as device configuration overhead and the potential security concerns of rooted devices. Such requirements are often undesirable for developers \cite{Lam:FSE17}. Conversely, our approach is able to work on any unmodified Android version without the necessity of a rooted device, requiring just a screen recording.

Nurmuradov \etal \cite{Nurmuradov:ISSTA17} introduced a record and replay tool for Android applications that captures user interactions by displaying the device screen in a web browser. %
This technique uses event data captured during the recording process to generate a heatmap that facilitate developers' understanding on how users are interacting with an application. This approach is limited in that users must interact with a virtual Android device through a web application, which could result in unnatural usage patterns. This technique is more focused towards session-based usability testing, whereas \vts is focused upon replaying "in-field" app usages from users or crowdsourced testers collected from real devices via screen recordings.

Tuovenen \etal \cite{Tuovenen} applied record and replay techniques to Android game application testing. They present MAuto, a tool that records screenshots to summarize user application interactions in the field, creates a test script for the objects and events that take place in the user interaction, and then replays the test. This allows mobile developers to improve their test suites without manually having to define or run these test cases. There are many similarities between MAuto and \vts, including the general execution flow of each pipeline. However, \vts is an end-to-end tool, while MAuto relies on AKAZE features to accomplish image-based object detection on the input screenshots, and then relies on the Appium framework to replay the events. Additionally, \vts requires less cooperation from the user by only requiring them to submit a screen-recording with the touch-indicator enabled on the device to show user taps.

Other work has focused on capturing high-level interactions in order to replay events \cite{Halpern:ISPASS15, airtest, culebra, espresso}. For instance Mosaic \cite{Halpern:ISPASS15}, uses an intermediate representation of user interactions to make replays device agnostic. %
Additional tools including HiroMacro \cite{hiromacro} and Barista \cite{Fazzini:ICST17} are Android applications that allow for a user to record and replay interactions. They require the installation or inclusion of underlying frameworks such as \texttt{replaykit} \cite{replaykit}, AirTest \cite{airtest}, or \texttt{troyd} \cite{Jeon:12}. Android Bot Maker \cite{bot_maker} is an Android application that allows for the automation of individual device actions, however, it does not allow for recording high-level interactions, instead one must enter manually the type of action and raw $(x,y)$ coordinates. In contrast to these techniques, one of \vtss primary aims is to create an Android record and replay solution which an inherently low barrier to usage. For instance, there are no frameworks to install, or instrumentation to add, the only input is a collectable screen recording. This makes \vts suitable for use in crowd- or beta-testing scenarios, and improves the likelihood of its adoption among developers for automated testing, given its ease of use relative to developer's perceptions of other tools~\cite{Linares-Vasquez:ICSME'17}.

	Finally, as crowdsourcing information from mobile app users has become more common with the advent of a number of testing services~\cite{applause,testbirds,mycrowd}, researchers have turned to utilizing usage data recorded from crowd-testers to enhance techniques related to automated test case generation for mobile apps. Linares-V\'{a}squez~\etal first introduced the notion of recording crowd-sourced data to enhance input generation strategies for mobile testing through the introduction of the {\sc MonkeyLab} framework~\cite{Linares-Vasquez:MSR'15}. This technique adapted N-gram language models trained on recorded user data to predict event sequences for automated mobile test case generation. Mao~\etal developed the {\sc Polariz} approach which is able to infer ``motif'' event sequences collected from the crowd that are applicable across different apps~\cite{Mao:ASE17}. The authors found that the activity-based coverage of the {\sc Sapienz} automated testing tool can be improved through a combination of crowd-based and search-based techniques. The two approaches discussed above require either instrumented or rooted devices ({\sc MonkeyLab}), or interaction with apps via a web-browser ({\sc Polariz}). Thus, \vts is complementary to these techniques as it provides a frictionless mechanism by which developers can collect user data in the form of videos.

\NEWEST{\subsection{Hybrid application testing}}

\NEWEST{Application testing is a crucial step in the mobile development process. Despite hybrid applications becoming more prevalent in the Google Play store ~\cite{Malavolta}, testing tools and techniques often ignore hybrid apps due to the challenges that they pose. Wan \etal ~\cite{Wan} discuss that because hybrid applications are built using web standards rather than traditional Android UI elements, app crawlers are not able to effectively test a hybrid application's entire functionality. Thus, other testing methods which rely on visual information can help to augment app crawling techniques. \vts provides an alternative means of effectively generating test scenarios for hybrid apps, using only visual information contained in videos. In the future, scenarios captured using \vts could be used to build models of hybrid apps for new types of hybrid app UI crawlers.}

\NEWEST{Issues related to hybrid apps have also been highlighted when exploring other testing challenges, such as test migration. Talebipour \etal ~\cite{Talebipour} explore UI test migration across mobile platforms (\eg Android or iOS). This paper illustrates that migrating tests between different platforms is more challenging than between different Android devices or versions, due to different UI design languages and icon usage. These challenges are similar to those that hybrid applications face as cross-platform apps are often built using frameworks that rely on web technologies such as React Native. Lin \etal ~\cite{Lin} develop {\sc CraftDroid}, a technique that transfers testing oracles between applications by tying GUI elements together based on their semantic functionality. This tool is not able to function with hybrid applications because it is unable to interpret the web elements that they use to accomplish functionality.}

\NEWEST{There has been a limited amount of past work that supports testing hybrid applications. Most notably, YazdaniBanafsheDaragh \etal ~\cite{YazdaniBanafsheDaragh} present \textit{Deep GUI}, a black-box testing framework that is meant to work well with applications that are built using a variety of implementations (\eg native and hybrid applications). The authors sought to build upon previous testing tools by learning from interactions based on the pixels of an application screen. \vts is similar to \textit{Deep GUI} in that it is a cross-platform approach, but it supports a different goal of recording specific user-provided test scenarios as opposed to generating input sequences based on a learned model. }

%% file: 8_threats.tex
\section{Limitations \& Threats to Validity}
\label{sec:limitations}
\vspace{0.2cm}
\noindent\textbf{Limitations.} 
Our approach has various limitations that serve as motivation for future work. One current limitation of the \Faster implementation our approach utilizes is that it is tied to the screen size of a single device, and thus a separate model must be trained for each screen size to which \approach is applied. However, as described in \secref{subsec:approach-detection}, the training data process is fully automated and models can be trained once and used for any device with a given screen size. This limitation could be mitigated by increasing dataset size including all type of screen sizes with a trade-off on the training time. To facilitate the use of our model by the research community, we have released our trained models for the two popular screen sizes of the Nexus 5 and Nexus 6P in our online appendix~\cite{appendix}.

Another limitation, which we will address in future work, is that our replayable traces are currently tied to the dimensions of a particular screen, and are not easily human readable. However, combining \approach with existing techniques for device-agnostic test case translation~\cite{Fazzini:ICST17}, and GUI analysis techniques for generating natural language reports~\cite{Moran:ICST16} could mitigate these limitations. 

As discussed in \secref{subsec:results-rq3}, the ability of \vts to faithfully replay swipes is also limited by the video frame-rate. 
During the evaluation, our devices were limited to 30fps, which made it difficult to completely resolve a small subset of gesture actions that were performed very quickly. However, this limitation could be addressed by improved Android hardware or software capable of recording video at or above 60fps, which, in our experience, should be enough to resolve nearly all rapid user gestures.

\NEW{Finally, \approach is also limited in that it performs noticeably better when utilized on a physical device rather than an emulated device. As discussed in \secref{subsec:results-rq3_baseline}, the unnatural way that users must interact with emulated devices and varying computer processing speeds produced poor input videos that were not well reproduced by \vts. We anticipate that conducting emulator screen recordings at a higher fps rate ($>$30 FPS) or on a machine with a faster or more powerful processor or graphical card might yield better results.}

\noindent\textbf{Internal Validity.} In our experiments evaluating \vts, threats to internal validity may arise from our manual validation of the correctness of replayed videos. To mitigate any potential subjectivity or errors, we had  two authors manually verify the correctness of the replayed scenarios. %
Furthermore, we have released all of our experimental data and code~\cite{appendix}, to facilitate the reproducibility of our experiments.

\NEW{\noindent\textbf{Construct Validity.} The main threat to construct validity arises from the potential bias in our manual creation of videos for the popular apps studies carried out to answer RQ$_3$ and RQ$_4$. It is possible that the author's knowledge of \vts influenced the manner in which we recorded videos.  To help mitigate this threat, we took care to record videos as naturally as possible (\eg normal speed, included natural quick gestures). Furthermore, we carried out an orthogonal study in the course of answering RQ$_3$, where users unfamiliar with \approach naturally recorded videos on physical and emulated devices, representing an unbiased set of videos. }

Another potential confounding factor concerns the quality of the dataset of screens used to train, test, and evaluate \vtss~\Faster and Opacity CNN. To mitigate this threat, we utilize the \ReDraw dataset~\cite{Moran:TSE18} of screens which have undergone several filtering and quality control mechanisms to ensure a diverse set of real GUIs. 
One more potential threat concerns our methodology for assessing the utility of \vts. Our developer interviews only assess the \textit{perceived} usefulness of our technique, determining whether developers actually receive benefit from \vts is left for future work.

\noindent\textbf{External Validity.} \NEW{Threats to the generalizability of our conclusions are mainly related to: (i) the number and diversity apps used in our evaluation; (ii) the representativeness of usage scenarios depicted in our experimental videos; and (iii) the generalizability of the responses given by the interviewed developers. To help mitigate the first threat, we performed 2 large-scale studies, 1 with 64 of the top native applications on Google Play mined from 32 categories, and 1 with 14 of the most popular hybrid applications from the Google Play store derived from 7 categories. While performing additional experiments with more applications is ideal, our experimental set of applications represents a reasonably large number of apps with different functionalities, which illustrate the relative applicability of \vts. To mitigate the second threat, we collected scenarios illustrating bugs, natural apps usages, real crashes, and controlled crashes from 8 participants. In our popular applications studies, we also gathered a variety of scenarios involving combinations of single-fingered and multi-fingered actions in sequence in an attempt to generalize natural app usages.} Finally, we do not claim that the feedback we received from developers generalizes broadly across industrial teams. However, the positive feedback and suggestions for future work we received in our interviews illustrate the potential practical usefulness of \vts. 

%% file: 9_conclusion.tex
\section{Conclusion \& Future Work}
\label{sec:conclusion}

\NEW{\approach is an approach for automatically translating video recordings of Android app usages into replayable scenarios.  A comprehensive evaluation reveals that \approach: (i) accurately identifies touch indicators and it is able to differentiate between opacity levels, (ii) is capable of reproducing a high percentage of complete scenarios in \textit{hybrid} and \textit{native applications} involving  \textit{single-} and \textit{multi-fingered actions} on  \textit{physical} and  \textit{emulated devices}, and (iii) is potentially useful to support real developers during a variety of tasks. }

By making use of screen recordings, which are notoriously difficult to analyze, \vts is the first automated translation tool of its kind. With all of our findings, we demonstrate that \vts may prove to be a invaluable tool for improving the rapid debugging processes that are inherent to mobile development. We expect developers may also be heavily inclined to incorporate it in place of other techniques due to \vtss low usage barrier, since \vts requires only a screen recording as input, and its modular structure, which can be easily manipulated to fit a wide variety of research and development tasks.

Our future work will focus on adding multiple improvements to \approach, including (i) producing scripts with coordinate-agnostic actions, (ii) generating natural language user scenarios, (iii) improving user experience via behavior analysis, and (iv) facilitating additional maintenance tasks via GUI-based information.